\documentclass[reqno,12pt]{amsart}
\usepackage[top=1.5in,right=1in,left=1in,bottom=1.5in]{geometry}
\usepackage{amssymb}
\usepackage{mathtools}
\usepackage{color}
\usepackage{tikz}
\usepackage{soul}
\usepackage[mathscr]{euscript}
\usepackage{float}

\usetikzlibrary{positioning,decorations.pathmorphing}
\usetikzlibrary{positioning,decorations.pathmorphing}
\definecolor{red}{rgb}{0.7,0,0}
\definecolor{grey}{RGB}{112,112,112}
\definecolor{blue}{RGB}{034,113,179}
\usepackage[colorlinks=true,citecolor=blue,linkcolor=red,urlcolor=grey,backref=page]{hyperref}

\newtheorem{theo}{Theorem}[section] 
\newtheorem{prop}[theo]{Proposition}  
\newtheorem{lemma}[theo]{Lemma}

\theoremstyle{remark}

\newcounter{mnotecount}[section]
\renewcommand{\star}{*}

\renewcommand{\themnotecount}{\thesection.\arabic{mnotecount}}

\newcommand{\mnote}[1]
{\protect{\stepcounter{mnotecount}}$^{\mbox{\footnotesize
$
\bullet$\themnotecount}}$ \marginpar{
\raggedright\tiny\em
$\!\!\!\!\!\!\,\bullet$\themnotecount: #1} }

\newcommand{\CP}{\mathbb{CP}}

\newcommand{\C}{\mathbb{C}}

\newcommand{\M}{\mathsf{M}}
\newcommand{\R}{\mathbb{R}}

\newcommand{\G}{\mathscr{G}}
\newcommand{\J}{\mathscr{J}}
\newcommand{\B}{\mathscr{B}}
\renewcommand{\S}{\mathscr{S}}
\renewcommand{\L}{\mathscr{L}}
\newcommand{\nvec}{\boldsymbol{n}}
\newcommand{\xivec}{\boldsymbol{\xi}}
\newcommand{\phivec}{\boldsymbol{\phi}}

\def\be{\begin{equation}}

\def\ee{\end{equation}}

\def\bea{\begin{eqnarray}}
\def\eea{\end{eqnarray}}

\numberwithin{equation}{section}
\begin{document} \date{\today}
\title{Shape modes of \(\mathbb{C}P^1\) vortices}

\author{Nora Gavrea$^1$, Derek Harland$^2$ and Martin Speight$^3$}
\address{School of Mathematics, University of Leeds, Leeds LS2 9JT, UK}
\email{N.A.Gavrea@leeds.ac.uk$^1$, D.G.Harland@leeds.ac.uk$^2$, J.M.Speight@leeds.ac.uk$^3$}

\begin{abstract}
In this paper we investigate the existence of internal modes of vortices in the gauged \(\mathbb{C}P^1\) sigma model. We develop a clean geometric formalism that highlights the symmetries of the Jacobi operator, obtained from the second variation of the energy functional. The formalism and subsequent results fundamentally rely on the Bogomol'nyi decomposition of the energy functional, and can therefore be extended to other models with such a decomposition. We prove the existence of at least one shape mode for a general \(\mathbb{C}P^1\) vortex solution on \(\mathbb{R}^2\), and find numerically the shape modes and corresponding frequencies of a radially symmetric vortex. A surprising result is that the shape mode eigenvalues are very close to the scattering threshold, suggesting weakly bound shape modes could be characteristic of the \(\mathbb{C}P^1\) model.
\end{abstract}

\maketitle
\section{Introduction}

The gauged $\mathbb{C}P^1$ sigma model (also known as the gauged $O(3)$ sigma model) was originally developed by Schroers \cite{schroers}.  It breaks the scale invariance of the pure $\mathbb{C}P^1$ sigma model by gauging a $U(1)$ symmetry. After including a suitable potential term, the model admits soliton solutions which provide minima of the associated energy functional within their homotopy class. They are conventionally called BPS as they arise as solutions of a first order PDE system obtained from a Bogomol'nyi-type argument \cite{schroers,CP1geometry}. This model and the abelian Higgs model have some similarities, for example the $U(1)$ symmetry of the fields and the fact that their solutions carry a quantized magnetic flux which makes them topologically stable. Solitons in both models are referred to as vortices.  

In the context of the abelian Higgs model, vortices have been studied from different perspectives and have become relevant in the study of both cosmology and superconductivity. In cosmology, they describe large-scale cosmic strings which lead to structure formation due to local concentrations of energy \cite{Shellard}. In the Ginzburg-Landau theory of superconductivity, they describe magnetic flux tubes occurring in a type-II superconducting material after it is cooled down below its critical temperature and a magnetic field is applied \cite{Abrikosov}. Similar applications have been discovered for the \(\mathbb{C}P^1\) vortices. In \cite{Yang_coexistence}, Yang has shown how to construct \(\mathbb{C}P^1\) vortices and anti-vortices coupled to gravity, representing cosmic strings and anti-strings with opposite magnetic charges. The \(\mathbb{C}P^1\) model has also been proposed to model the competition between superconducting phases, due to its target space \(\mathbb{C}P^1\cong S^2\) which comes with an additional constraint, allowing competition between a superconducting component and a charge density wave  component \cite{intervortexforces,CP1_superconductivity}.

The gauged $\C P^1$ model has two interesting new features with no counterpart in the abelian Higgs model. First it carries an extra parameter $\tau\in(-1,1)$, the height of the vacuum manifold on $S^2$. (The analogous parameter in the abelian Higgs model can be scaled away.)  This parameter affects the size and energy of vortices nontrivially. Second,
due to its compact target space, the \(\mathbb{C}P^1\) model exhibits two different types of vortices. Both these features were first noticed by Kimm, Lee and Lee \cite{kimleelee}, in the context of the model with a Chern Simons term and additional neutral scalar field. In general, vortex positions are the points in physical space $\Sigma$ that are mapped by the Higgs field $\phi$ to a fixed point of the action of the gauge group $U(1)$ on the target space. In the abelian Higgs model, the target space $\C$ has only one fixed point, the origin, so there is only one species of vortex. In the gauged $\C P^1$ model, the target space $S^2$ has two fixed points which we may take, without loss of generality, to be the $\pm\nvec=(0,0,\pm1)$, hence two distinct species of vortex. These may generically coexist. Finite energy field configurations carry a pair of integer valued topological invariants, the numbers $k_+$, $k_-$ of preimages of these fixed points counted with orientation and multiplicity. Every static energy minimizer in the $(k_+,k_-)$ class defines a disjoint pair of effective divisors on $\Sigma$ of degrees $k_+$ and $k_-$ consisting of the preimages $D_+=\phi^{-1}(\nvec)$ and $D_-=\phi^{-1}(-\nvec)$. $D_+\cap D_-=\emptyset$ since no point can be mapped simultaneously to both $\nvec$ and $-\nvec$. Conversely, given a disjoint pair of effective divisors $D_+,D_-$ of degrees $(k_+,k_-)$ there is a static vortex in the degree $(k_+,k_-)$ class, unique up to gauge, with $\phi^{-1}(\pm\nvec)=D_\pm$. This was proved on $\Sigma=\R^2$ by Yang
in the symmetric case $\tau=0$ \cite{Yang_multisolitons_existence} and Han for general $\tau$ in
\cite{Han_general_taubes_existence}. On a compact Riemann surface $\Sigma$ (of sufficiently large area) this was proved by Sibner, Sibner and Yang in the special case $\tau=0$ in \cite{Yang_coexistence_compact}, and in the case of general $\tau$ by Garcia Lara \cite{Garcia-Lara_coexistence_compact}.  The space of all these $(k_+,k_-)$ vortex solutions modulo gauge equivalence is called the moduli space, which is a generically non-compact complex manifold of dimension
$k_++k_-$. The moduli space has a natural Riemannian metric whose geodesic flow models the dynamics of slowly moving vortices. A detailed analysis of the geometry of the moduli space of \(\mathbb{C}P^1\) vortices is given in \cite{CP1geometry}.

Our focus in this paper is the perturbative modes around a BPS vortex solution, along with the frequency spectrum of these modes. They solve an eigenvalue problem of a certain elliptic operator, arising from the second variation of the energy functional of the model. Quantum mechanically, they can be interpreted as particle states: they can be either bound states (particles trapped in the core of the vortex) or unbound states (particles scattering with the vortex). Bound states have $L^2$ normalizable eigenfunctions, while scattering states do not. By a shape mode, we will mean a bound state with strictly positive eigenvalue. 
(Bound modes with eigenvalue $0$ correspond to perturbations tangent to the space of static solutions; all such modes are generated by moving the vortex positions, or by gauge transformations.) 

In classical dynamics, shape modes correspond to spatially localized oscillatory motions of the time dependent field equations linearized about the static vortex. Their existence, frequency and degeneracy are known to have profound effects on the low energy scattering of abelian Higgs vortices 
\cite{shapemode_scattering_AH_2026,spectralflow,collectivecoord,Krusch_Rees_Winyard_scattering,modulispace_excited_vortex_2025}
and they have been intensively investigated in that model
\cite{alberto_note_on_bound_states,dissecting,alberto_AHshape_modes_2025} by numerical means.  But to date the are no rigorous analytic existence results for shape modes, either in the abelian Higgs model or the gauged $\mathbb{C}P^1$ model.

In this paper, we investigate the existence of shape modes of vortices in the \(\mathbb{C}P^1\) model, focusing on the case \(\Sigma=\mathbb{R}^2\). Rather than follow the supersymmetry argument developed in \cite{dissecting} for radially symmetric abelian Higgs vortices, we develop a new formalism which allows for a cleaner treatment of the second variation of the energy functional as well as the eigenvalue problem giving rise to the internal modes, working initially on a general surface \(\Sigma\), and later specializing to \(\mathbb{R}^2\). This formalism fundamentally exploits the Bogomol'nyi energy decomposition, hence in principle it can be applied to any model with this property. It can be easily applied to vortices in the abelian Higgs model with slight modifications, and it could also be applied to other types of topological solitons (such as monopoles or lumps). We present this formalism and its implications in Section 3, where we prove
the existence of at least one shape mode for a general \(\mathbb{C}P^1\) vortex solution on \(\Sigma=\mathbb{R}^2\), which is the main result of this paper. In Section 4, we focus on the radially symmetric case with vortices coincident at the origin and use our previous results to compute the shape modes numerically. We find that for a charge one vortex, the shape mode eigenvalue is very close to the scattering threshold, suggesting that bound states in the \(\mathbb{C}P^1\) model are only weakly bound.

\section{\texorpdfstring{The gauged \(\mathbb{C}\text{P}^1\) sigma model}{The gauged CP1 sigma model}}
\subsection{The energy functional}
Let \(\Sigma\) be an oriented Riemannian surface, and \(P\to\Sigma\) a principal \(U(1)\)-bundle equipped with a connexion \(A\). Let \(\boldsymbol{\phi}\) be a section of the associated bundle \(P\times_{\rho}M\to\Sigma\), where \(M\) is the target manifold, and \(\rho\) is a Hamiltonian action of \(U(1)\) on \(M\).

In this paper we consider \(\Sigma=\mathbb{R}^2\) and \(M=\mathbb{C}P^1\) which we identify as \(S^2\) with the round metric. The action of \(U(1)\) on \(M=S^2\) is represented by rotations about the \(z\)-axis.     We will often adhere to the notation \(\Sigma\), before we specialize our results to the particular case \(\Sigma=\mathbb{R}^2\). In this later case, \(P\) is trivial and the Higgs field is a function \(\boldsymbol{\phi}:\mathbb{R}^2\to S^2\).

We are interested in finding pairs \((A,\boldsymbol{\phi})\) which are global minimizers of the energy functional
\begin{equation}\label{energyfunctional}
E[\boldsymbol{\phi}, A]=\dfrac{1}{2}\int_{\Sigma}^{}\left[\left|F\right|^2+\left|D\boldsymbol{\phi}\right|^2+\left|\tau-\boldsymbol{n}\cdot\boldsymbol{\phi}\right|^2\right], 
\end{equation}
where \(\boldsymbol{n}=(0,0,1)\) is the constant unit vector pointing in the \(z\)-direction, \(F=dA\) is the curvature of the connexion, and \(D\boldsymbol{\phi}=d\boldsymbol{\phi}-A\boldsymbol{n}\times\boldsymbol{\phi}\) is the covariant derivative in a local trivialization of $P\times_\rho M$. 
The parameter \(\tau\in(-1,1)\) in the potential \(U(\boldsymbol{\phi})=(\tau-\boldsymbol{n}\cdot\boldsymbol{\phi})^2\) defines the vacuum manifold, where the potential attains its minimum value equal to zero.  (The case \(\tau=\pm 1\) is somewhat pathological, and will not be considered here.)

For the non-compact case \(\Sigma=\mathbb{R}^2\), finite energy solutions require the following boundary conditions:
\begin{equation}\label{boundaryconditions}
\lim_{|\boldsymbol{x}|\to\infty}F=0, \text{ } \lim_{|\boldsymbol{x}|\to\infty}D\boldsymbol{\phi}=0, \text{ } \lim_{|\boldsymbol{x}|\to\infty}\boldsymbol{n}\cdot\boldsymbol{\phi}=\tau.
\end{equation}

\subsection{Gauge transformations}
The energy functional (\ref{energyfunctional}) is invariant under the following local gauge transformations
\[\boldsymbol{\phi}\mapsto R(\theta)\boldsymbol{\phi}\]
\[A\mapsto A+d\theta, \]
where \(\theta:\Sigma\to\mathbb{R}\) and \(R(\theta)\in SO(3)\) represents the rotation with angle \(\theta\) about the \(z\)-axis
\[R(\theta)=\begin{pmatrix}
\cos{\theta} & -\sin{\theta} & 0\\
\sin{\theta} & \cos{\theta} & 0 \\
0 & 0 & 1
\end{pmatrix}.\]

\subsection{The Bogomol'nyi equations}
The energy functional admits a Bogomol'nyi-type decomposition, which results from completing the square inside the energy functional:

\begin{equation}\label{energy_bog_decomposition}
E[\boldsymbol{\phi}, A]=\dfrac{1}{2}\int_{\Sigma}^{}\left[\dfrac{1}{2}\left|D\boldsymbol{\phi}- \star\boldsymbol{\phi}\times D\boldsymbol{\phi}\right|^2+\left|\star F+ (\tau-\boldsymbol{n}\cdot\boldsymbol{\phi})\right|^2\right]
\end{equation}
\[+\int_{\Sigma}^{}\left[\dfrac{1}{2}D\boldsymbol{\phi}\cdot \boldsymbol{\phi}\times D\boldsymbol{\phi}- (\tau-\boldsymbol{n}\cdot\boldsymbol{\phi})F\right].\]

It can be shown \cite{CP1geometry} that the cross term integrates to 
\[\int_{\Sigma}^{}\left[\dfrac{1}{2}D\boldsymbol{\phi}\cdot \boldsymbol{\phi}\times D\boldsymbol{\phi}- (\tau-\boldsymbol{n}\cdot\boldsymbol{\phi})F\right]=2\pi(1-\tau)k_++2\pi(1+\tau)k_-,\]
under the assumption that \(F\), \(D\boldsymbol{\phi}\) and \(\boldsymbol{n}\cdot\boldsymbol{\phi}\) decay sufficiently rapidly on the boundary of \(\Sigma\). 

The numbers \(k_+\) and \(k_-\) are integers encoding the number of North vortices and South anti-vortices on \(\Sigma\), respectively: the pre-image of the North pole under \(\boldsymbol{\phi}\) is a set of \(k_+\) points on \(\Sigma\), representing the locations of the North vortices, and the pre-image of the South pole under \(\boldsymbol{\phi}\) is a set of \(k_-\) points on \(\Sigma\), representing the locations of the South anti-vortices. These (anti)-vortices are topologically distinct, and their co-existence was established by Yang \cite{Yang_coexistence}. The difference between vortices and anti-vortices arises from the fact that local coordinate systems in the neighbourhoods of the North and South poles have opposite orientations. Thus, the field winds around the vacuum manifold \(k_+\) times in anticlockwise direction, corresponding to the contribution of the North vortices, and \(k_-\) times in clockwise direction, corresponding to the contribution of the South anti-vortices. These windings are also related to the total magnetic flux through the surface \cite{intervortexforces}, which is quantized and can be computed
\[\dfrac{1}{2\pi}\int_{\Sigma}F=k_+-k_-.\]

From the above Bogomol'nyi argument, we conclude the energy is bounded below by the topological invariant given by the integration of the cross terms, and the inequality is saturated when the first order Bogomol'nyi equations hold
\begin{equation} \label{B1}
D\boldsymbol{\phi}=\star\boldsymbol{\phi}\times D\boldsymbol{\phi}
\end{equation}
\begin{equation}  \label{B2}
\star F=-(\tau-\boldsymbol{n}\cdot\boldsymbol{\phi}),
\end{equation}
with solutions \((A,\boldsymbol{\phi})\) being conventionally called BPS solutions. It has been shown that, up to gauge equivalence, BPS configurations are determined by the pair of divisors \((D_+,D_-)\) with degrees \((k_+,k_-)\) on \(\Sigma\), specifying the locations of the vortices and anti-vortices, respectively, including their algebraic multiplicities \cite{CP1geometry,Yang_multisolitons_existence}.

\section{\texorpdfstring{Shape modes of a \(\mathbb{C}\text{P}^1\) vortex solution}{Shape modes of a CP1 vortex solution}}
Recall the shape modes represent the perturbative modes around a BPS vortex solution. They are eigenmodes of the second order elliptic operator obtained from the second variation of the energy functional (\ref{energyfunctional}), with positive eigenvalue. They offer extensive information about the dynamics of excited vortices near the critical BPS point in the moduli space. Such dynamics is rich and complicated, and has been widely studied recently in the context of the Abelian-Higgs model \cite{spectralflow,dynamics_excited_3vortices,collectivecoord,Krusch_Rees_Winyard_scattering}, revealing surprising phenomena, such as for example the existence of spectral walls \cite{spectral_wall}.

To investigate shape modes in the context of the \(\mathbb{C}P^1\) model, we must first find the second variation of the energy functional (\ref{energyfunctional}) around a BPS solution \((\boldsymbol{\phi},A)\). It is helpful to work initially on a general surface \(\Sigma\), and later set \(\Sigma=\mathbb{R}^2\). Throughout we assume that \(P\) is trivial. We will start by computing the first variation and defining some new operators, as this establishes a good starting point for the second variation whose calculation is more computationally difficult.

\subsection{The first variation.}
Let \((\boldsymbol{\phi}_s,A_s)\) be a 1-parameter variation of the BPS solution 
\((\boldsymbol{\phi}_0,A_0)=(\boldsymbol{\phi},A)\). 
This generates the infinitesimal perturbation
\[\left.\dfrac{d}{ds}\right\vert_{s=0}\boldsymbol{\phi}_s(\boldsymbol{x})=\boldsymbol{\eta}(\boldsymbol{x}), \text{ } \left.\dfrac{d}{ds}\right\vert_{s=0}A_s(\boldsymbol{x})=\alpha(\boldsymbol{x}).\]

We will consider the energy functional written in Bogomol'nyi form, which will later enable us to write the second variation operator as a composition of two first order operators. This approach becomes very helpful once we try to prove existence of and construct the shape modes. We have the energy 
\[E[\boldsymbol{\phi}, A]=\dfrac{1}{4}\|D\boldsymbol{\phi}-\star\boldsymbol{\phi}\times D\boldsymbol{\phi}\|_{L^2}^2+\dfrac{1}{2}\|\star F+(\tau-\boldsymbol{n}\cdot\boldsymbol{\phi})\|^2_{L^2}+E_0,\]
where \(E_0=2\pi(1-\tau)k_++2\pi(1+\tau)k_-\), and the \(L^2\) inner product is given by
\[\left<\alpha,\beta\right>_{L^2}=\int_{\Sigma}^{}\alpha\wedge\star\beta.\]
By defining the operator
\begin{align*}Bog: 
C^\infty\left(\Sigma, S^2\right)\oplus\Gamma{(T^\star\Sigma)}&\to\Gamma{(T^\star\Sigma\otimes\boldsymbol{\phi}^\star TS^2)}\oplus C^\infty(\Sigma) \\
\begin{pmatrix}
\boldsymbol{\phi} \\
A
\end{pmatrix}&\mapsto 
\begin{pmatrix}
\dfrac{1}{\sqrt{2}}\left(D\boldsymbol{\phi}-\star\boldsymbol{\phi}\times D\boldsymbol{\phi}\right) \\[15pt]
\star F+(\tau-\boldsymbol{n}\cdot\boldsymbol{\phi})
\end{pmatrix},\end{align*}
we can re-write the energy more compactly as
\begin{equation} \label{energy_compactform}
E[\boldsymbol{\phi}, A]=\dfrac{1}{2}\left<Bog(\boldsymbol{\phi},A), Bog(\boldsymbol{\phi}, A)\right>_{L^2}+E_0.
\end{equation}
Then the first variation of the energy is given by 
\[\left.\dfrac{d}{ds}\right|_{s=0}E[\boldsymbol{\phi}_s,A_s]=\left<\left.\dfrac{d}{ds}\right|_{s=0}Bog(\boldsymbol{\phi}_s,A_s), Bog(\boldsymbol{\phi}_s, A_s)\right>_{L^2}=\left<\mathscr{B}\boldsymbol{\xi}, Bog(\boldsymbol{\phi}, A)\right>_{L^2},\]
where \(\boldsymbol{\xi}=\begin{pmatrix} \boldsymbol{\eta}\\
\alpha
\end{pmatrix}\) and $\mathscr{B}$ is the differential of the map $Bog$ at $(\phi,A)$:
\begin{align*}\mathscr{B}:\Gamma(\boldsymbol{\phi}^*TS^2)\oplus \Gamma(T^{\star}\Sigma)&\to\Gamma(T^{\star}\Sigma\otimes\boldsymbol{\phi}^*TS^2)\oplus C^{\infty}(\Sigma)\\
\boldsymbol{\xi}=\begin{pmatrix}\boldsymbol{\eta}\\
\alpha\end{pmatrix}&\mapsto 
\begin{pmatrix}
\dfrac{1}{\sqrt{2}}\left(D\boldsymbol{\eta}-\alpha\boldsymbol{n}\times\boldsymbol{\phi}-\star\boldsymbol{\eta}\times D\boldsymbol{\phi}-\star\boldsymbol{\phi}\times (D\boldsymbol{\eta}-\alpha\boldsymbol{n}\times\boldsymbol{\phi})\right) \\[15pt]
\star d\alpha-\boldsymbol{n}\cdot\boldsymbol{\eta}
\end{pmatrix}. \end{align*}
The domain of this operator will occur frequently in the sequel, so it is useful to give it a name. Since it is the space of linear perturbations about the static vortex $(\boldsymbol{\phi},A)$  we call it
\begin{equation}
    \mathcal{PERT}:=\Gamma(\boldsymbol{\phi}^\star TS^2)\oplus\Gamma(T^*\Sigma)
\end{equation}

To rewrite the operator \(\mathscr{B}\) more compactly and to ensure an easier calculation for the second variation in the next subsection, it is helpful to define the following linear operators
\begin{eqnarray}
L_{\boldsymbol{\phi}}:
\Gamma(T^{\star}\Sigma\otimes\boldsymbol{\phi}^{*} TS^2)
&\to&
\Gamma(T^{\star}\Sigma\otimes\boldsymbol{\phi}^{*} TS^2)\nonumber 
\\
\boldsymbol{\eta}
&\mapsto&
\frac{1}{\sqrt{2}}\left(\boldsymbol{\eta}-\boldsymbol{\phi}\times\star\boldsymbol{\eta}\right)
\label{linear_map}
\\
P_{\boldsymbol{\phi}}:\Gamma(T^{*}\Sigma\otimes\mathbb{R}^3)&\to&\Gamma(T^{\star}\Sigma\otimes\boldsymbol{\phi}^{*} TS^2)\nonumber \\
\boldsymbol{u}&\mapsto& \boldsymbol{u}-(\boldsymbol{u}\cdot\boldsymbol{\phi})\boldsymbol{\phi}.
\label{projection_map}
\end{eqnarray}
One easily verifies that
\begin{align}
L_{\boldsymbol{\phi}}\circ L_{\boldsymbol{\phi}}&=\sqrt{2}L_{\boldsymbol{\phi}}\label{prop1}\\
P_{\boldsymbol{\phi}}\circ P_{\boldsymbol{\phi}}&=P_{\boldsymbol{\phi}}\label{prop2}\\
\left[L_{\boldsymbol{\phi}},\star\right]&=0\label{prop3}\\
\left[L_{\boldsymbol{\phi}},P_{\boldsymbol{\phi}}\right]&=0\label{prop4}\\
L_{\boldsymbol{\phi}}^{\dagger}&=L_{\boldsymbol{\phi}}\label{prop5}\\
P_{\boldsymbol{\phi}}^{\dagger}&=P_{\boldsymbol{\phi}},\label{prop6}
\end{align}
where the ``dagger" superscript denotes the \(L^2\)-adjoint of the respective operator, and
\begin{equation}\label{operator_B}
\mathscr{B}\begin{pmatrix}\boldsymbol{\eta}\\
\alpha\end{pmatrix}=
\begin{pmatrix}
L_{\boldsymbol{\phi}}\circ P_{\boldsymbol{\phi}}\left(D\boldsymbol{\eta}-\alpha\boldsymbol{n}\times\boldsymbol{\phi}\right) \\[5pt]
\star d\alpha-\boldsymbol{n}\cdot\boldsymbol{\eta}
\end{pmatrix}.\end{equation}

\subsection{The second variation.}
Let \((\boldsymbol{\phi}_{s,t},A_{s,t})\) be a 2-parameter variation of the BPS solution 
\((\boldsymbol{\phi}_{0,0},A_{0,0})=(\boldsymbol{\phi},A)\). This generates the infinitesimal perturbations
\[\left.\dfrac{\partial}{\partial s}\right\vert_{s=0}\boldsymbol{\phi}_{s,t}(\boldsymbol{x})=\boldsymbol{\eta}_t(\boldsymbol{x}), \text{ } \left.\dfrac{\partial}{\partial s}\right\vert_{s=0}A_{s,t}(\boldsymbol{x})=\alpha_t(\boldsymbol{x})\]
\[\left.\dfrac{\partial}{\partial t}\right\vert_{s=t=0}\boldsymbol{\phi}_{s,t}(\boldsymbol{x})=\tilde{\boldsymbol{\eta}}(\boldsymbol{x}), \text{ } \left.\dfrac{\partial}{\partial t}\right\vert_{s=t=0}A_{s,t}(\boldsymbol{x})=\tilde{\alpha}(\boldsymbol{x}).\]
Let $\boldsymbol{\xi}=(\boldsymbol{\eta}_0,\alpha_0)$ and
$\tilde{\boldsymbol{\xi}}=(\tilde{\boldsymbol{\eta}},\tilde\alpha)$

The second variation of the energy functional around a BPS solution is then given by
\begin{eqnarray} 
\left.\dfrac{\partial}{\partial t}\right\vert_{t=0}\left.\dfrac{\partial}{\partial s}\right\vert_{s=0}E[\boldsymbol{\phi}_{s,t}, A_{s,t}]&=&\left<\left.\dfrac{\partial}{\partial s}\right|_{s=0}Bog(\boldsymbol{\phi}_{s,t},A_{s,t}), \left.\dfrac{\partial}{\partial t}\right|_{t=0}Bog(\boldsymbol{\phi}_{s,t}, A_{s,t})\right>_{L^2}\nonumber \\
&=&\left<\mathscr{B}\boldsymbol{\xi}, \mathscr{B}\tilde{\boldsymbol{\xi}}\right>_{L^2}\nonumber \\
&=&\left<\boldsymbol{\xi}, \mathscr{B}^{\dagger}\mathscr{B}\tilde{\boldsymbol{\xi}}\right>_{L^2},
\label{second_variation1}
\end{eqnarray}
where
\begin{equation}\label{operator_B_dagger}
\begin{aligned}\mathscr{B}^{\dagger}: \text{ }\Gamma(T^{\star}\Sigma\otimes\boldsymbol{\phi}^*TS^2)\oplus C^{\infty}(\Sigma)&\to\mathcal{PERT}\\
 \begin{pmatrix}\boldsymbol{\chi_1}\\
\chi_2\end{pmatrix}&\mapsto 
\begin{pmatrix}
P_{\boldsymbol{\phi}}\left(-\star D\star L_{\boldsymbol{\phi}}(\boldsymbol{\chi_1})-\boldsymbol{n}\chi_2\right)\\[5pt]
-\star d\chi_2-(\boldsymbol{n}\times\boldsymbol{\phi})\cdot L_{\boldsymbol{\phi}}\left(\boldsymbol{\chi}_1\right)
\end{pmatrix} \end{aligned}
\end{equation}
is the \(L^2\)-adjoint of \(\mathscr{B}\) expressed in (\ref{operator_B}). Its calculation follows quickly by using the properties listed in (\ref{prop1})-(\ref{prop6}).

Recall that the perturbation \(\boldsymbol{\eta}\) comes with the constraint \(\boldsymbol{\eta}(\boldsymbol{x})\in T_{\boldsymbol{\phi}(\boldsymbol{x})}S^2\), and hence we should be careful with handling inner products of the type \(\left<\boldsymbol{\eta}, \cdot\right>_{L^2}\) when deriving expression (\ref{operator_B_dagger}), since only the terms perpendicular to \(\boldsymbol{\phi}\) bring a contribution, as the ones parallel to \(\boldsymbol{\phi}\) equate to zero when taken in the inner product with \(\boldsymbol{\eta}\). To rule out the occurence of such extraneous terms, we note \(\left<\boldsymbol{\eta}, \cdot\right>_{L^2}=\left<\boldsymbol{\eta},P_{\boldsymbol{\phi}}(\cdot)\right>_{L^2}\).

Note from (\ref{second_variation1}) we can now identify the second variation operator \(\mathscr{J}\), called the Jacobi operator, with \(\mathscr{J}=\mathscr{B}^{\dagger}\mathscr{B}\). This decomposition reinforces more obviously that \(\mathscr{J}\) is a positive semi-definite self-adjoint operator.

An important observation is that we were able to decompose the Jacobi operator into a product of two first order operators, one of which is the \(L^2\)-adjoint of the other, simply due to the fact that the energy functional admits a Bogomol'nyi argument. Hence this represents a general property of this type of models.

\subsection{Zero modes and the orthogonal gauge condition}
Recall the energy functional (\ref{energyfunctional}) is invariant under the gauge transformations presented in subsection 2.2. Therefore, perturbations of the fields represented by infinitesimal gauge transformations leave the Bogomol'nyi equations invariant, and lie in the kernel of the operator \(\mathscr{B}\).

Let us denote by $\mathscr{G}$ the map which assigns to a smooth real function on $\Sigma$ the infinitesimal gauge transformation that it generates:
\begin{eqnarray} \label{G_operator}
\mathscr{G}:C^\infty(\Sigma)&\to&\mathcal{PERT}\nonumber \\
\chi&\mapsto&(\chi\boldsymbol{n}\times\boldsymbol{\phi},d\chi).
\end{eqnarray}
As we have just observed, $\B\G=0$, that is, infinitesimal gauge transformations lie in the kernel of $\B$. If we wish to contruct the tangent space to the vortex moduli space at (the gauge equivalence class of) $(\phivec,A)$, we should keep only those vectors in $\ker \B$ which are $L^2$ orthogonal to all such gauge transformations, i.e.\ $\langle \mathscr{G}(\chi),(\boldsymbol{\eta},\alpha)\rangle_{L^2}=0$ for all $\chi$. Clearly this is equivalent to
\begin{equation}\label{GC}
    \mathscr{G}^\dagger(\boldsymbol{\eta},\alpha)=0
\end{equation}
where
\begin{eqnarray}\label{G_dagger_operator}
\mathscr{G}^\dagger:\mathcal{PERT}&\to&C^\infty(\Sigma)\nonumber \\
(\boldsymbol{\eta},\alpha)&\mapsto& -\star d \star\alpha +\boldsymbol{\eta}\cdot\left(\boldsymbol{n}\times\boldsymbol{\phi}\right)
\end{eqnarray}
is the $L^2$ adjoint of $\mathscr{G}$. 

Hence $T_{[(\phivec,A)]}\M$, the tangent space of the moduli space \(\M\) at the point \(([\boldsymbol{\phi},A)]\), is precisely the kernel of the extended operator
\begin{equation}\label{B^G_operator}
\mathscr{B}^G\coloneq\mathscr{B}\oplus\mathscr{G}^\dagger:
\mathcal{PERT}\to\mathcal{BOG} 
\end{equation}
whose codomain we have denoted
\[
\mathcal{BOG}:=\Gamma(T^{\star}\Sigma\otimes\boldsymbol{\phi}^*TS^2)\oplus C^{\infty}(\Sigma)\oplus C^{\infty}(\Sigma).
\]
The \(L^2\)-adjoint of \(\mathscr{B}^G\) is
\begin{equation}\label{B^G_dagger_operator}
\begin{aligned}
\mathscr{B}^{G\dagger}: \mathcal{BOG} &\to \mathcal{PERT}\\
\begin{pmatrix}
\boldsymbol{\chi}_1 \\
\chi_2\\
\chi_3
\end{pmatrix}&\mapsto\begin{pmatrix}
P_{\boldsymbol{\phi}}\left(-\star D\star L_{\boldsymbol{\phi}}(\boldsymbol{\chi_1})-\boldsymbol{n}\chi_2\right)+\boldsymbol{n}\times\boldsymbol{\phi}\chi_3\\[5pt]
-\star d\chi_2-(\boldsymbol{n}\times\boldsymbol{\phi})\cdot L_{\boldsymbol{\phi}}\left(\boldsymbol{\chi}_1\right)+d\chi_3
\end{pmatrix},
\end{aligned}
\end{equation}
using the properties (\ref{prop1})-(\ref{prop6}).

Since $\J=\B^\dagger\B$, it follows immediately that
\begin{equation}\label{JG=0}
\J\G=0,\qquad \G^\dagger\J=0.
\end{equation}
Hence, every eigenmode of $\J$ not in its kernel  automatically satisfies the gauge orthogonality condition \eqref{GC}.

\subsection{The extended Jacobi operator}
We can now define the extended Jacobi operator by
\begin{equation} \label{new_Jacobi}
\mathscr{J}^G\coloneq\mathscr{B}^{G\dagger}\mathscr{B}^G=\mathscr{B}^{\dagger}\mathscr{B}+\mathscr{G}\mathscr{G}^{\dagger}=\mathscr{J}+\mathscr{G}\mathscr{G}^{\dagger}:\mathcal{PERT}\to\mathcal{PERT}.
\end{equation}
We will prove shortly that $\mathscr{J}$ and $\mathscr{J}^G$ have the same eigenvalue spectrum.

Analyzing the structure of the operators \(\mathscr{B}^G\) and \(\mathscr{B}^{G\dagger}\), it is helpful to define the $L^2$ isometries
\begin{equation}\label{S1}
\begin{aligned}
\mathscr{S}_1:\mathcal{PERT} &\to \mathcal{PERT} \\
\begin{pmatrix}
\boldsymbol{\eta}\\
\alpha
\end{pmatrix} &\mapsto
\begin{pmatrix}
\boldsymbol{\phi}\times\boldsymbol{\eta}\\
\star\alpha
\end{pmatrix}
\end{aligned}
\end{equation}

\begin{equation}\label{S2}
\begin{aligned}
\mathscr{S}_2:\mathcal{BOG}&\to \mathcal{BOG} \\
\begin{pmatrix}
\boldsymbol{\chi}_1\\
\chi_2\\
\chi_3
\end{pmatrix} &\mapsto
\begin{pmatrix}
\star\boldsymbol{\chi}_1\\
-\chi_3\\
\chi_2
\end{pmatrix}.
\end{aligned}
\end{equation}
The following lemma concerns the properties of the maps (\ref{S1}) and (\ref{S2}) and the symmetries of the operators \(\mathscr{J}^G\) and \(\mathscr{B}^{G}\mathscr{B}^{G\dagger}\).

\lemma{The maps \(\mathscr{S}_1\) and \(\mathscr{S}_2\) satisfy the following properties
\begin{align}
\mathscr{S}_1\circ \mathscr{S}_1&=-id_{\mathcal{PERT}} \label{S1_prop1}\\
\mathscr{S}_2\circ \mathscr{S}_2&=-id_{\mathcal{BOG}}\label{S2_prop1}\\
\mathscr{S}_1^{\dagger}&=-\mathscr{S}_1\label{S1_prop2}\\
\mathscr{S}_2^{\dagger}&=-\mathscr{S}_2\label{S2_prop2}\\
\mathscr{G}^{\dagger}\mathscr{S}_1\mathscr{G}&=0\label{S1G_prop}
\end{align}
Furthermore, \(\left[\mathscr{J}^G,\mathscr{S}_1\right]=0\) and \(\left[\mathscr{B}^G\mathscr{B}^{G\dagger},\mathscr{S}_2\right]=0\), i.e. they represent symmetries of the operators \(\mathscr{J}^G\) and \(\mathscr{B}^{G}\mathscr{B}^{G\dagger}\), respectively.}\label{jacobi_sym} \\\\
\normalfont
\textit{Proof}. From the definitions of the two maps given in (\ref{S1}) and (\ref{S2}) it is immediate to check that the properties (\ref{S1_prop1})-(\ref{S2_prop2}) hold. 

To prove (\ref{S1G_prop}) we compute directly by using the definitions (\ref{G_operator}), (\ref{G_dagger_operator}), (\ref{S1})

\[\mathscr{G}^{\dagger}\mathscr{S}_1\mathscr{G}\chi=\mathscr{G}^{\dagger}\begin{pmatrix}
\boldsymbol{\phi}\times(\boldsymbol{n}\times\boldsymbol{\phi})\chi\\
\star d\chi
\end{pmatrix}=-\star d\star \left(\star d\chi\right)+\left[\boldsymbol{\phi}\times(\boldsymbol{n}\times\boldsymbol{\phi})\chi\right]\cdot (\boldsymbol{n}\times\boldsymbol{\phi})=0.\]
From the expressions (\ref{B^G_operator}) and (\ref{B^G_dagger_operator}) we can easily check that
\begin{equation} \label{sym1}
\mathscr{B}^G\mathscr{S}_1=\mathscr{S}_2\mathscr{B}^G
\end{equation}
\begin{equation} \label{sym2}
\mathscr{B}^{G\dagger}\mathscr{S}_2=\mathscr{S}_1\mathscr{B}^{G\dagger},
\end{equation}
where the second equation can also be obtained by taking the \(L^2\)-adjoint of the first one.

Applying \(\mathscr{B}^{G\dagger}\) to \eqref{sym1} we obtain
\[\mathscr{B}^{G\dagger}\mathscr{B}^G\mathscr{S}_1=\mathscr{B}^{G\dagger}\mathscr{S}_2\mathscr{B}^G=\mathscr{S}_1\mathscr{B}^{G\dagger}\mathscr{B}^G,\]
where in the second equality we used (\ref{sym2}). Hence the operators \(\mathscr{J}^G\) and \(\mathscr{S}_1\) commute
\begin{equation} \label{J_sym}
\mathscr{J}^G\mathscr{S}_1=\mathscr{S}_1\mathscr{J}^G,
\end{equation}
which is equivalent to saying that \(\mathscr{S}_1\) is a symmetry of the operator \(\mathscr{J}^G\). Similarly, applying $\mathscr{B}^G$ to \eqref{sym2} yields 
\begin{equation} \label{the_other_sym}
\mathscr{B}^{G}\mathscr{B}^{G\dagger}\mathscr{S}_2=\mathscr{S}_2\mathscr{B}^{G}\mathscr{B}^{G\dagger}.
\end{equation}
\qedsymbol\\\\
\textit{Remarks.}
\begin{enumerate}
\item The map $\mathscr{S}_1$ is an isometry of $\mathcal{PERT}$ (with respect to its natural $L^2$ inner product) which squares to $-id$. Hence it may naturally be interpreted as an almost complex structure on $\mathcal{PERT}$. Its restriction to $\ker \mathscr{B}^G$ is precisely the natural almost complex structure on the vortex moduli space.  
\item The above Lemma implies that if \(\boldsymbol{\xi}\) is an eigenmode of \(\mathscr{J}^{G}\) with eigenvalue \(\lambda^2\), then so is \(\mathscr{S}_1\boldsymbol{\xi}\), and similarly if \(\boldsymbol{\chi}\) is an eigenmode of \(\mathscr{B}^{G}\mathscr{B}^{G\dagger}\) with eigenvalue \(\lambda^2\), then so is \(\mathscr{S}_2\boldsymbol{\chi}\). So eigenmodes of $\mathscr{J}^G$ come in degenerate pairs. 
\item We also notice that if \(\boldsymbol{\chi}\) is an eigenmode of \(\mathscr{B}^{G}\mathscr{B}^{G\dagger}\) with eigenvalue \(\lambda^2\), then \(\mathscr{B}^{G\dagger}\boldsymbol{\chi}\) (if non-zero) is an eigenmode of \(\mathscr{J}^{G}\) with the same eigenvalue. This observation will turn out to be crucial later when we compute the shape modes, as the operator \(\mathscr{B}^{G}\mathscr{B}^{G\dagger}\) allows for the decoupling of the PDE system obtained in the eigenvalue problem.
\item Finally, observe that if \(\boldsymbol{\xi}\in \ker  \mathscr{J}^G\), then 
\[0=\left<\boldsymbol{\xi},\mathscr{J}^G\boldsymbol{\xi}\right>_{L^2}=\left<\mathscr{B}^{G}\boldsymbol{\xi},\mathscr{B}^{G}\boldsymbol{\xi}\right>_{L^2}=\left|\mathscr{B}^{G}\boldsymbol{\xi}\right|^2_{L^2},\]
implying \(\mathscr{B}^{G}\boldsymbol{\xi}=0\). Hence  \(\ker\mathscr{J}^G=\ker\mathscr{B}^G\), and similarly \(\ker\mathscr{B}^{G}\mathscr{B}^{G\dagger}=\ker\mathscr{B}^{G\dagger}\). 
\end{enumerate}

Now that we are equipped with these new operators, we show that this formulation allows for an elegant and quick computation of a vortex shape mode. 

\theo{If \(\psi:\Sigma\to\mathbb{R}\) is an \(L^2\) eigenfunction of the Schr\"odinger operator
\begin{equation} \label{PDE}
L=\Delta+|\boldsymbol{n}\times\boldsymbol{\phi}|^2
\end{equation}
 then \(\mathscr{S}_1\mathscr{G}\psi\) is an \(L^2\)-integrable eigenmode of \(\mathscr{J}\) with the same eigenvalue. Here $\Delta=-\star d\star d$ denotes the Laplace-Beltrami operator on $\Sigma$.}
\label{theorem_GGdagger}\\\\
\normalfont
\textit{Proof}. 
First observe that $L=\mathscr{G}^{\dagger}\mathscr{G}$. Applying \(\mathscr{G}\) to the left, we observe that if \(L\psi=\lambda^2\psi\), then 
\[
\mathscr{G}\G^\dagger(\mathscr{G}\psi)=\lambda^2\G\psi
\]
and $\ker\G=0$ by its definition \eqref{G_operator}, so $\G\psi$
is an eigenmode of \(\mathscr{G}\mathscr{G}^{\dagger}\) with the same eigenvalue. Now $\J\G=0$ since, for all $\chi\in C^\infty(\Sigma)$, $\G(\chi)$ is an infinitesimal gauge transformation, all of which lie in the kernel of $\J$ by the gauge invariance of the energy functional. Hence
\[
\mathscr{J}^G\mathscr{G}\psi=(\mathscr{J}+\mathscr{G}\mathscr{G}^{\dagger})\mathscr{G}\psi=\mathscr{G}\mathscr{G}^{\dagger}\mathscr{G}\psi=\lambda^2\mathscr{G}\psi,\]
that is, \(\mathscr{G}\psi\) is an eigenmode of \(\mathscr{J}^G\) with eigenvalue \(\lambda^2\).

By the symmetry we proved in Lemma \ref{jacobi_sym}, we deduce that \(\mathscr{S}_1\mathscr{G}\psi\) is also an eigenmode of \(\mathscr{J}^G\). Then we have 
\[
\mathscr{J}\mathscr{S}_1\mathscr{G}\psi=(\mathscr{J}^G-\mathscr{G}\mathscr{G}^\dagger)\mathscr{S}_1\mathscr{G}\psi=
\mathscr{J}^G\left(\mathscr{S}_1\mathscr{G}\psi\right)=\lambda^2\mathscr{S}_1\mathscr{G}\psi,\]
by \eqref{S1G_prop} proved in Lemma \ref{jacobi_sym}. We therefore deduce that \(\mathscr{S}_1\mathscr{G}\psi\) is also an eigenmode of \(\mathscr{J}\). 

The \(L^2\)-integrability part of the theorem follows quickly by noticing

\[|\mathscr{S}_1\mathscr{G}\psi|^2_{L^2}=\left<\mathscr{S}_1\mathscr{G}\psi,\mathscr{S}_1\mathscr{G}\psi\right>_{L^2}=\left<\psi,\mathscr{G}^{\dagger}\mathscr{S}_1^{\dagger}\mathscr{S}_1\mathscr{G}\psi\right>_{L^2}=\lambda^2\left<\psi,\psi\right>_{L^2}<\infty,\]
where we used (\ref{S1_prop2}), (\ref{S1_prop1}) and \(\mathscr{G}^{\dagger}\mathscr{G}\psi=\lambda^2\psi\). Hence if \(\psi\) is \(L^2\)-integrable, then so is \(\mathscr{S}_1\mathscr{G}\psi\). \qedsymbol \\

\textit{Remarks.} 
\begin{enumerate}
\item Since the potential $|\nvec\times\phivec|^2$ in the Schr\"odinger operator is non-negative and vanishes only on the divisors $D_\pm$, any $L^2$ eigenmode constructed in this way must have $\lambda^2>0$, and so is a shape mode.

\item The above proposition reveals a powerful and quick method to construct a shape mode, by solving a single scalar PDE $L\psi=\lambda^2\psi$. This is much simpler than solving the eigenvalue problem for $\J$ directly (a coupled system of 5 PDEs with a constraint).  However,
there is no guarantee that this method will give all possible shape modes. 
\end{enumerate}

To find all possible shape modes of a given vortex solution, we need to solve the eigenvalue problem for the operator \(\mathscr{J}\), which fails to be elliptic. We next show that the eigenvalue spectrum of $\J$ coincides precisely with that of $\J^G$, which {\em is} elliptic. 
The proof will make use of the orthogonal decomposition
\[
\mathcal{PERT}=\mathcal{G}_\infty\oplus\mathcal{G}_\infty^\perp
\]
where $\mathcal{G}_\infty=\text{Im}\G$ is the space of infinitesimal gauge transformations and
$\mathcal{G}_\infty^\perp=\ker\G^\dagger$ is its $L^2$ orthogonal complement.

\prop{Assume \(k_+\neq0\)  or \(k_-\neq0\). Then the operators \(\mathscr{J}\) and \(\mathscr{J}^G=\mathscr{J}+\mathscr{G}\mathscr{G}^{\dagger}\) have the same \(L^2\)-eigenvalues, corresponding to eigenmodes which are \(L^2\)-integrable.}\label{jacobi_spectrum}\\\\
\normalfont
\textit{Proof}. We first observe that $0$ is an eigenvalue of both $\J$ and $\J^G$ since any tangent vector to $\M$ at $[(\phivec,A)]$ is in the kernel of both $\B$ and $\B^G$. It remains to show that $\J,\J^G$ have the same non-zero eigenvalues.

First assume $\xivec\neq 0$ is $L^2$ and $\J\xivec=\lambda^2\xivec$ with $\lambda^2\neq 0$. Then it follows from \eqref{JG=0} that $\G^\dagger\xivec=0$ and hence
$\J^G\xivec=\J\xivec=\lambda^2\xivec$. Hence $\xivec$ is an $L^2$ eigenmode of $\J^G$ with the same eigenvalue. 

Conversely, assume that $\xivec\neq 0$ is $L^2$ and $\J^G\xivec=\lambda^2\xivec$ with $\lambda^2\neq 0$.
Denote by
\[\boldsymbol{\xi}=\boldsymbol{\xi}^g+\boldsymbol{\xi}^{g\perp},\]\
where \(\boldsymbol{\xi}^g\in\mathcal{G}_{\infty}\), \(\boldsymbol{\xi}^{g\perp}\in\mathcal{G}_{\infty}^{\perp}\) its decomposition into $\mathcal{G}_\infty\oplus\mathcal{G}_\infty^\perp$. Then we have, by \eqref{JG=0}, 
\[
\mathscr{J}^G\boldsymbol{\xi}=\mathscr{J}\boldsymbol{\xi}^{g\perp}+\mathscr{G}\mathscr{G}^{\dagger}\boldsymbol{\xi}^g=\lambda^2(\boldsymbol{\xi}^g+\boldsymbol{\xi}^{g\perp}),\]
whose components in $\mathcal{G}_\infty^\perp$ and $\mathcal{G}_\infty$ are
\begin{equation} \label{G_eval_problem}
\mathscr{G}\mathscr{G}^{\dagger}\boldsymbol{\xi}^g=\lambda^2\boldsymbol{\xi}^g
\end{equation}
\begin{equation} \label{J_eval_problem}
\mathscr{J}\boldsymbol{\xi}^{g\perp}=\lambda^2\boldsymbol{\xi}^{g\perp}
\end{equation}
respectively. 
Hence, if $\xivec^{g\perp}\neq0$, it is an $L^2$ eigenmode of $\J$, and the claim is established.

Finally, consider the case that $\xivec^{g\perp}=0$, so $\xivec=\xivec^g=\G(\chi)$ for some $\chi\in C^\infty(\Sigma)$. Then $\G^\dagger \S_1\xivec^g=\G^\dagger \S_1\G(\chi)=0$ 
by equation (\ref{S1G_prop}), so
\[
\J \S_1\xivec^g=\J^G S_1\xivec^g=\S_1\J^G\xivec^g=\lambda^2\S_1\xivec^g,
\]
by Lemma \ref{jacobi_sym}. $\S_1$ is an $L^2$ isometry, so $\S_1\xivec^g$ is an $L^2$ eigenmode of $\J$ with eigenvalue $\lambda^2$, which completes the proof.   
\qedsymbol
\\

Our next existence result builds on Theorem \ref{theorem_GGdagger}.

\theo{Let \((\boldsymbol{\phi}, A)\) be a solution to the Bogomol'nyi equations on \(\Sigma=\mathbb{R}^2\) with nonempty pair of divisors \((D_+,D_-)\). Then there exists a bound state of the BPS \(\mathbb{C}P^1\) vortex if either of the following holds
\begin{itemize}
\item \(k_-=|D_-|=0\) and \(\tau\in(0,1)\)
\item \(k_+=|D_+|=0\) and \(\tau\in(-1,0)\)
\item \(\tau=0\).
\end{itemize}}\label{theorem_shapemodes_existence}
\normalfont
\vspace{0.2cm}

\textit{Proof}. In Theorem \ref{theorem_GGdagger}, we proved that if \(\psi\) is an eigenmode of \(\mathscr{G}^{\dagger}\mathscr{G}\), then \(\mathscr{S}_1\mathscr{G}\psi\) is an eigenmode of \(\mathscr{J}\). Hence if we can find a bound state of the Schr\"odinger operator (\ref{PDE}), then \(\mathscr{S}_1\mathscr{G}\psi\) represents a bound state of the vortex solution.

We can re-write the eigenvalue problem for (\ref{PDE}) as
\begin{equation} \label{Schrodinger_PDE}
-\nabla^2\psi+V\psi=E\psi,
\end{equation}
where \(E=\lambda^2-(1-\tau^2)\) and \(V=\left|\boldsymbol{n}\times\boldsymbol{\phi}\right|^2-(1-\tau^2)=\tau^2-(\boldsymbol{n}\cdot\boldsymbol{\phi})^2\). We observe that this new potential has the property that
\[\lim_{|\boldsymbol{x}|\to\infty}V(\boldsymbol{x})=\lim_{|\boldsymbol{x}|\to\infty}[\tau^2-(\boldsymbol{n}\cdot\boldsymbol{\phi})^2]=0,\]
due to the boundary conditions on \(\boldsymbol{\phi}\). We now use a maximum principle to prove that \(V(\boldsymbol{x})\leq0\) everywhere on the plane in the cases specified. 

In order to do so, we first project stereographically from the South pole of \(S^2\). This allows us to instead work with the complex field \(u=\dfrac{\phi_1+i\phi_2}{1+\phi_3}\). The Bogomol'nyi equations then transform to 
\[\tilde{D}_1u=-i\tilde{D}_2u\]
\[\star dA=-\left(\tau-\dfrac{1-|u|^2}{1+|u|^2}\right),\]
where now the covariant derivative is \(\tilde{D}u=du-iAu\).
Note we can write 
\begin{equation} \label{mod_u}
|u|^2=\dfrac{1-\phi_3}{1+\phi_3},
\end{equation}
and define 
\begin{equation} \label{v_def}
v=\ln{\dfrac{1-\phi_3}{1+\phi_3}}+\ln{\dfrac{1+\tau}{1-\tau}}. 
\end{equation}
Using the boundary conditions on \(\boldsymbol{\phi}\), we have \(\lim_{|\boldsymbol{x}|\to\infty}v(\boldsymbol{x})=0\). Note that $D_+,D_-$ are the collections of points in $\R^2$ at which $\phivec=\nvec$ and $-\nvec$ respectively, so at each $\boldsymbol{p}\in D_+$, \(v(\boldsymbol{p})=-\infty\) and at each 
$\boldsymbol{q}\in D_-$, \(v(\boldsymbol{q})=\infty\).

Combining the two Bogomol'nyi equations, we obtain the so called Taubes equation

\begin{equation} \label{Taubes}
\nabla^2v=2\dfrac{e^v-1}{\dfrac{1}{1-\tau}+\dfrac{e^v}{1+\tau}}+4\pi\sum_{\boldsymbol{q}\in D_-}\delta(\boldsymbol{q})-4\pi\sum_{\boldsymbol{p}\in D_+}\delta(\boldsymbol{p}).
\end{equation}
In \cite{Yang_multisolitons_existence}, this equation was derived for the case \(\tau=1\). In \cite{Han_general_taubes_existence}, it was proven that (\ref{Taubes}), as a particular example of a more general Taubes-type equation considered by the author, admits a unique solution for any parameter \(\tau\in(-1,1)\), that this solution is smooth away from $D_\pm$ and exponentially localized. 

We observe that \(\phi_3\geq\tau\) is equivalent to \(v\leq0\). We now assume \(k_-=0\), so \(v\) only has negative singularities, and \(\tau\in(0,1)\). If \(v\) is not everywhere non-positive, then since \(\lim_{|\boldsymbol{x}|\to\infty}v(\boldsymbol{x})=0\), it must achieve a maximum at some \(z_0\), such that \(v(z_0)>0\). At this maximum, the Hessian \(Hess(v(z_0))\) is negative semidefinite, hence \(Tr(Hess(v(z_0))\leq 0\), but the trace of the Hessian is equal  \(\nabla^2 v\), and therefore \(\nabla^2v(z_0)\leq 0\). Observe that the LHS of (\ref{Taubes}) is non-positive at \(z_0\), while the RHS is positive, leading to a contradiction. We deduce that \(\phi_3(\boldsymbol{x})\geq\tau\) everywhere on the plane. Since \(\tau>0\), then we also have \((\phi_3)^2\geq\tau^2\), and hence \(V(\boldsymbol{x})\leq0\) everywhere on the plane.

The case \(k_+=0\) and \(\tau\in(-1,0)\) follows from a similar argument, where now the signs are flipped and we obtain \(\phi_3\leq\tau\) everywhere on the plane. Since \(\tau<0\), we reach the same conclusion that \((\phi_3)^2\geq\tau^2\), and thus again \(V(\boldsymbol{x})\leq0\). 

The last case follows from the simple observation that \(\tau=0\) implies immediately that the potential \(V(\boldsymbol{x})=\tau^2-(\phi_3)^2\) is everywhere non-positive without any assumption on the number of vortices and anti-vortices.

A simple variational method with a cleverly chosen test function shows that any smooth potential \(V(\boldsymbol{x})\) in two dimensions obeying \(V(\boldsymbol{x})\leq0\) for all \(\boldsymbol{x}\in\mathbb{R}^2\), $V(\boldsymbol{x})<0$ at some $\boldsymbol{x}$ and \(\lim_{|\boldsymbol{x}|\to\infty}V(\boldsymbol{x})=0\) supports at least one bound state \cite{boundstate_2D}. We note that for a bound state we must have \(E<0\), so the eigenvalue is restricted to \(\lambda^2<1-\tau^2\).
We therefore conclude that there exists at least one shape mode of the vortex solution for each of the three cases considered. \qedsymbol\\

We can improve Theorem \ref{theorem_shapemodes_existence} by extending the range of \(\tau\) for which at least one bound state exists by using an explicit test function and a continuity argument.

\theo{Let \((\boldsymbol{\phi}, A)\) be a solution of the Bogomol'nyi equations on \(\Sigma=\mathbb{R}^2\) with a fixed nonempty pair of divisors \((D_+,D_-)\). Then there exists a bound state of the BPS \(\mathbb{C}P^1\) vortex if any of the following holds
\begin{itemize}
\item \(k_-=|D_-|=0\) and \(\tau\in(-\tau_\star,1)\)
\item \(k_+=|D_+|=0\) and \(\tau\in(-1,\tau_\star)\)
\item \(\tau\in(-\tau_\star,\tau_\star)\),
\end{itemize}\label{theorem_shapemodes_existence_extended} where \(\tau_\star\in(0,1)\) depends on the divisor pair \((D_+,D_-)\).}
\normalfont
\vspace{0.2cm}

\textit{Proof}. In order to extend the range of \(\tau\), we need to construct a test function and apply the variational method.

Taking inner products on both sides and integrating by parts in (\ref{Schrodinger_PDE}) we can express the energy as the Rayleigh-Ritz quotient

\begin{equation} \label{Rayleigh_Ritz}
E=\dfrac{\left<\psi,H\psi\right>_{L^2}}{\left<\psi,\psi\right>_{L^2}}=\dfrac{1}{||\psi||^2_{L^2}}\int_{\mathbb{R}^2}^{}\left(|d\psi|^2+V(r,\theta)\psi^2\right)rdrd\theta,
\end{equation}
where the Hamiltonian operator is given by \(H=L-(1-\tau^2)=-\nabla^2+V(r,\theta)\), and \((r,\theta)\) are polar coordinates on the plane.

Let \(E_0\) be the ground state energy of \(H\). The variational method states that the energy of an arbitrary state is always at least \(E_0\). Hence if we can find a state for which the Rayleigh quotient is negative, then we can infer that the true ground state energy \(E_0\) is also negative, which leads to the existence of at least one bound state.

Given \(\tau\), consider the family of test functions
\begin{equation} \label{approx_wf}
\psi_{\alpha,\tau}(r)=e^{-\sqrt{1-\tau^2-\alpha^2}r},
\end{equation}
with \(\alpha^2<1-\tau^2\).
Plugging into (\ref{Rayleigh_Ritz}) we obtain the associated energy 
\[E(\alpha,\tau)=\dfrac{2}{\pi}(1-\tau^2-\alpha^2)\int_{\mathbb{R}^2}^{}\left(1-\alpha^2-(\phi_3)^2\right)\psi_{\alpha,\tau}^2rdrd\theta,\]
where recall that the potential can be written as \(V=\tau^2-(\phi_3)^2\).
Let 
\begin{equation} \label{integral}
I_{\alpha}=\int_{\mathbb{R}^2}^{}\left(1-\alpha^2-(\phi_3)^2\right)\psi_{\alpha,\tau}^2rdrd\theta,
\end{equation}
and note that \(I_\alpha\) is finite for all \(\alpha^2<1-\tau^2\), since \(\lim_{|\boldsymbol{x}|\to\infty}\phi_3(\boldsymbol{x})=\tau\) and \(\psi_{\alpha,\tau}\) is exponentially decaying. Furthermore, if \(\alpha^2=1-\tau^2\), then
\[I_{\sqrt{1-\tau^2}}=\int_{\mathbb{R}^2}^{}\left(\tau^2-(\phi_3)^2\right)rdrd\theta\]
is also finite due to the exponential convergence of \(\phi_3\) to \(\tau\). 

Before continuing our argument for extending the range of \(\tau\) for which shape modes exist, we first need to establish that \(E(\alpha,\tau)\) is continuous in \(\tau\) in a small neighbourhood of \(\tau=0\). Notice that the energy depends on \(\phi_3\) which itself depends on $\tau$, and hence we need to establish some form of continuity of $\phi_3$ with respect to $\tau$. The following theorem, whose proof is presented in an appendix, suffices:

\theo{Given a disjoint pair of divisors (\(D_+,D_-\)), let \((A^\tau,\boldsymbol{\phi}^\tau)\) be the solution (unique up to gauge) of the Bogomol'nyi equations (\ref{B1}) and (\ref{B2}) with parameter \(\tau\). Then there exists $\epsilon\in(0,1)$ such that the map
\begin{align*}\Phi:(-\epsilon,\epsilon)&\to L^2(\mathbb{R}^2)\\
\tau&\mapsto \phi_3^\tau-\tau
\end{align*}
is \(C^1\).\label{thm_taubes_continuity}}
\normalfont
\vspace{0.2cm}

 In the remaining part of the proof we will refer to \(\phi_3\) by \(\phi_3^\tau\) to emphasize its dependence on \(\tau\).

A simple calculation shows that 
\[E(\alpha, \tau)=1-\alpha^2-\dfrac{2}{\pi}(1-\tau^2-\alpha^2)\int_{\mathbb{R}^2}^{}(\phi_3^\tau)^2\psi_{\alpha,\tau}^2rdrd\theta.\]
It therefore remains to prove that \(\tau\mapsto \phi_3^\tau\psi_{\alpha,\tau}\) is continuous in \(L^2\). Recalling \(\tau\mapsto\Phi(\tau)=\phi_3^\tau-\tau\) is continuous on \(\tau\) in a small neighbourhood of \(\tau=0\) by Theorem \ref{thm_taubes_continuity}, we can rewrite
\begin{eqnarray}
\int_{\mathbb{R}^2}(\phi_3^\tau)^2\psi_{\alpha,\tau}^2d^2x&=&\int_{\mathbb{R}^2}^{}(\phi_3^\tau-\tau)^2\psi_{\alpha,\tau}^2d^2x+2\tau\int_{\mathbb{R}^2}^{}\phi_3^\tau\psi_{\alpha,\tau}^2d^2x-\tau^2\int_{\mathbb{R}^2}^{}\psi_{\alpha,\tau}^2d^2x
\nonumber \\
&=&\int_{\mathbb{R}^2}^{}\Phi(\tau)^2\psi_{\alpha,\tau}^2d^2x+2\tau\int_{\mathbb{R}^2}^{}\Phi(\tau)\psi_{\alpha,\tau}^2d^2x+\tau^2\int_{\mathbb{R}^2}^{}\psi_{\alpha,\tau}^2d^2x
\end{eqnarray}
Note 
\[\tau^2\int_{\mathbb{R}^2}^{}\psi_{\alpha,\tau}^2d^2x=\dfrac{\tau^2\pi}{2(1-\tau^2-\alpha^2)}\]
is obviously continuous in \(\tau\) in a small neighbourhood of \(\tau=0\). Now we deal with the remaining two integrals. Consider \(\tau,\tau_0\in(-\epsilon,\epsilon)\), where \(0<\epsilon\ll1\). Then we have 
\begin{eqnarray}
&&\left|\int_{\mathbb{R}^2}\Phi(\tau)\psi_{\alpha,\tau}^2d^2x
-\int_{\mathbb{R}^2}\Phi(\tau_0)\psi_{\alpha,\tau_0}^2d^2x\right|\nonumber \\
&=&\left|\int_{\mathbb{R}^2}^{}(\Phi(\tau)-\Phi(\tau_0))\psi_{\alpha,\tau}^2d^2x+\int_{\mathbb{R}^2}^{}\Phi(\tau_0)(\psi_{\alpha,\tau}^2-\psi_{\alpha,\tau_0}^2)d^2x\right|\nonumber \\
&\leq& \left|\langle\Phi(\tau)-\Phi(\tau_0),\psi_{\alpha,\tau}^2\rangle_{L^2}\right|+\left|\langle\psi_{\alpha,\tau}-\psi_{\alpha,\tau_0},\Phi(\tau_0)(\psi_{\alpha,\tau}+\psi_{\alpha,\tau_0})\rangle_{L^2}\right|\nonumber \\
&\leq& \|\Phi(\tau)-\Phi(\tau_0)\|_{L^2}\|\psi_{\alpha,\tau}^2\|_{L^2}+\|\psi_{\alpha,\tau}-\psi_{\alpha,\tau_0}\|_{L^2}\|\Phi(\tau_0)(\psi_{\alpha,\tau}+\psi_{\alpha,\tau_0})\|_{L^2}\nonumber \\
&\to& 0 \text{ as }\tau\to\tau_0
\end{eqnarray}
since \(\Phi\) and \(\tau\mapsto\psi_{\alpha,\tau}\) are \(L^2-\)continuous in a small neighourhood of \(\tau=0\), and the other \(L^2-\)norms are finite due to their integrands' exponential decay.

In a very similar manner we also prove that \(\tau\mapsto\int_{\mathbb{R}^2}^{}\Phi(\tau)^2\psi_{\alpha,\tau}^2d^2x\) is continuous in a small neighbourhood around \(\tau=0\). This completes the proof that \(\tau\mapsto\int_{\mathbb{R}^2}(\phi_3^\tau)^2\psi_{\alpha,\tau}^2\) is continuous, and therefore we deduce that \(E(\alpha,\tau)\) is continuous on \(\tau\) in a small neighbourhood around \(\tau=0\). 

We can now use this fact to prove the statement of the theorem. Assume \(\tau=0\) and \(k_\pm\) are not both $0$. Then obviously \(I_1<0\). Since \(I_{\alpha}\) depends continuously on \(\alpha\) (the \(\alpha\) dependence of the integrand is explicit through polynomial and exponential functions), there exists \(\alpha_{\star}<1\) such that \(I_{\alpha_{\star}}<0\), and hence \(E(\alpha_{\star},0)<0\). Since \(E(\alpha_{\star},\tau)\) is continuous in \(\tau\), we deduce there exists a small neighbourhood \(\tau\in(-\tau_\star,\tau_\star)\) for which \(E(\alpha_{\star},\tau)<0\), and hence a bound state exists. Notice this argument depends on the solution \((\boldsymbol{\phi},A)\), and therefore \(\tau_\star\) depends on the locations of the (anti-)vortices, information which is encoded in the divisor pair \((D_+,D_-)\).

We can now easily extend this argument to the cases where \(k_-=0\) or \(k_+=0\). The above argument proved there exists a bound state for any \(k_\pm\) if \(\tau\in(-\tau_\star,\tau_\star)\), so in particular we can set \(k_-=0\). From Theorem (\ref{theorem_shapemodes_existence}), a bound state exists for \(k_-=0\) and \(\tau\in(0,1)\). Combining these two results we deduce there exists a bound state when \(k_-=0\) and \(\tau\in(-\tau_\star,1)\). Similarly we deduce we have a bound state when \(k_+=0\) and \(\tau\in(-1,\tau_\star)\). \qedsymbol

\section{Radially symmetric vortices}

In this section, we apply our previous results in the particular case where we assume radial symmetry and all vortices are concentrated at the origin. The Bogomol'nyi equations reduce to two coupled ODEs. We compute the shape modes numerically by solving the Bogomol'nyi equations and the ODE reduced from (\ref{Schrodinger_PDE}) together as a system.

\subsection{\texorpdfstring{Symmetry reduction on \(\Sigma=\mathbb{R}^2\).}{Symmetry reduction on Sigma = R2.}}
If we assume radial symmetry, the locations of all (anti)-vortices are constrained to the origin of \(\mathbb{R}^2\). Therefore, we can only have either \(N\) North vortices (\(k_+=N\), \(k_-=0\)) or \(N\) South anti-vortices (\(k_+=0\), \(k_-=N\)). Choosing the former case, as the latter follows in a very similar manner, we look for solutions using the hedgehog ansatz

\begin{equation} \label{higgs_hedgehog}
\boldsymbol{\phi}(x)=(\sin{f(r)}\cos{N\theta}, \sin{f(r)}\sin{N\theta}, \cos{f(r)})
\end{equation}

\begin{equation} \label{gauge_hedgehog}
A=Na(r)d\theta,
\end{equation}
where \((r,\theta)\) are polar coordinates on the plane.

Plugging these in the Bogomol'nyi equations, they reduce to a system of first order ODEs for \(a(r)\) and \(f(r)\):
\begin{equation} \label{RB1}
f'= \dfrac{N}{r}(1-a)\sin{f}   
\end{equation}
\begin{equation} \label{RB2}
a'=-\dfrac{r}{N}(\tau-\cos{f}).   
\end{equation}

The condition \(\boldsymbol{\phi}(0)=(0,0,1)\) that all vortices are superposed at the origin implies that \(f(0)=0\). Furthermore, regularity of the gauge field at the origin imposes \(a(0)=0\), and the boundary conditions we set before reduce to 
\[\lim_{r\to\infty}{a'(r)}=0 \text{, } \lim_{r\to\infty}a(r)=1  \text{, }\lim_{r\to\infty}{\cos{f(r)}}=\tau
.\]
Hence, as \(r\to\infty\), \(\boldsymbol{\phi}\) approaches the vacuum manifold given by 
\[(\sqrt{1-\tau^2}\cos{N\theta}, \sqrt{1-\tau^2}\sin{N\theta}, \tau).\]

\subsection{Asymptotic behaviour and magnetic flux}
By looking at the asymptotic behaviour of (\ref{RB1}) and (\ref{RB2}) in the two regimes \(r\to0\), \(r\to\infty\), we can find analytic approximations for the functions \(f(r)\), \(a(r)\). 
\begin{itemize}
    \item As \(r\to0\), we find

    \begin{equation}\label{small_r_asymp_a}
    a(r)\approx \dfrac{r^2}{2N}(1-\tau)
    \end{equation}
    \begin{equation}\label{small_r_asymp_f}
    f(r)\approx Ar^N,
    \end{equation}
    where \(A\) is an arbitrary constant, and will be used as a shooting constant to find the vortex solution numerically.
    \item As \(r\to\infty\), we find
    \begin{equation}\label{large_r_asymp_a}
    a(r)\approx 1-\dfrac{q}{2N\pi}rK_1(r\sqrt{1-\tau^2})
    \end{equation}
    \begin{equation}\label{large_r_asymp_f}
    f(r)\approx \arccos{\tau}-\dfrac{q}{2\pi}K_0(r\sqrt{1-\tau^2}),
    \end{equation}
    where \(K_0\), \(K_1\) are modified Bessel functions of the second kind. The constant \(q\) represents the strength of the asymptotic vortex charge in the particle intepretation of Speight \cite{Speight_particle_interpretation,intervortexforces}. These asymptotics hold in the case \(\tau\in(-1,1)\).
    
\end{itemize}

A quick calculation using the hedgehog ansatz shows that \(F=Na'(r)dr\wedge d\theta\). Hence  we can directly compute the total magnetic flux
\[\int_{\mathbb{R}^2}^{}F=2\pi N,\]
which is quantized since \(N\) is an integer. One should note that this result depends crucially on our choice to restrict $\tau$ to $(-1,1)$. If \(\tau=\pm1\),  the magnetic flux is not quantised, but it is equal to \(N\alpha\), where \(\alpha=\lim_{r\to\infty}{a(r)}\) 
\cite{schroers}. 

\subsection{\texorpdfstring{Shape modes of a cylindrically symmetric \(\mathbb{C}\text{P}^1\)\(N\)-vortex}{Shape mode solutions of a cylindrically symmetric CP1 N-vortex}} We follow the strategy given by Theorem \ref{theorem_GGdagger} which allows us to easily compute at least one shape mode of the vortex solution by firstly solving for the eigenfunctions of the Schr\(\ddot{\text{o}}\)dinger operator (\ref{PDE}). To prove that this strategy indeed gives us all the possible shape modes in the case of a radially symmetric \(N\)-vortex, we then construct the operator \(\mathscr{B}^{G}\mathscr{B}^{G\dagger}\) in radial coordinates, and show it reduces to the single PDE given by the eigenvalue problem of (\ref{PDE}).

In the case of radial symmetry, substituting the hedgehog ansatz (\ref{higgs_hedgehog}) shows that the operator (\ref{PDE}) becomes 
\[L=\Delta + \sin^2f(r).\]
To find the eigenfunctions of \(L\) we need to solve the eigenvalue problem \(L\Psi=\lambda^2\Psi\). Motivated by the \(U(1)\) symmetry, we make the ansatz \(\Psi(r,\theta)=\psi(r)e^{ik\theta}\), where \(k\) is an integer. The eigenvalue problem then reduces to 
\begin{equation} \label{radial_schrodinger}
-\psi''-\dfrac{1}{r}\psi'+\left(\dfrac{k^2}{r^2}+\sin^2{f}\right)\psi=\lambda^2\psi,
\end{equation}
Since the potential depends explicitly on the radial profile of the Higgs field, we need to solve (\ref{radial_schrodinger}) together with (\ref{RB1}) and (\ref{RB2}). Theorem \ref{theorem_GGdagger} showed that the shape mode is then given by
\[\mathscr{S}_1\mathscr{G} \Psi(r)=\begin{pmatrix}
(\boldsymbol{n}-(\boldsymbol{n}\cdot\boldsymbol{\phi})\boldsymbol{\phi})\psi(r)e^{ik\theta}\\
r\psi'(r)e^{ik\theta}d\theta-i\dfrac{k}{r}\psi e^{ik\theta}dr
\end{pmatrix}=\begin{pmatrix}
\boldsymbol{\eta}\\
\alpha
\end{pmatrix}.\]
and can be visualised in Figures \ref{profiles_N=1}, \ref{profiles_N=2} and \ref{profiles_N=3} for the cases \(N=1,2,3\), and \(k=0\), where we define 
\begin{equation}\label{def_psi1}
    \psi_1(r)\coloneq|\eta_3|=(1-(\phi_3)^2)\psi(r)
\end{equation}
\begin{equation}\label{def_psi2}
    \psi_2(r)\coloneq r\psi'(r).
\end{equation}

\subsection{Shape mode asymptotics}
By looking at the asymptotic behaviour of the Schr\(\ddot{\text{o}}\)dinger equation (\ref{radial_schrodinger}) in the two regimes \(r\to0\) and \(r\to\infty\), we can find analytic approximations for the bound state solutions \(\psi(r)\). 
\begin{itemize}
    \item As \(r\to0\), we find
    \begin{equation}\label{small_r_asymp_p}
    \psi(r)\approx C_0 J_k(\lambda r)
    \end{equation}
    where \(J_k\) is a Bessel function of the first kind and \(C_0\) is an arbitrary constant. The ODE is linear and hence \(C_0\) can be chosen arbitrarily. In particular, we can always rescale it such that \(C_0=1\), which we will use in the numerical calculations. 
    \item As \(r\to\infty\), we find
    \begin{equation}\label{large_r_asymp_p}
    \psi(r)\approx CK_k(\sqrt{1-\tau^2-\lambda^2}r)
    \end{equation}
    where \(K_k\) is a modified Bessel function of the second kind and \(C\) is a constant. These asymptotics hold in the case \(\tau\in(-1,1)\). Notice we obtain bound states if and only if \(\lambda^2<1-\tau^2\).
\end{itemize}

\subsection{Numerical results}
We solve numerically for the shape modes in the radially symmetric case on \(\mathbb{R}^2\), where the vortices are all located at the origin. Since the eigenvalue problem for the Jacobi operator involves the Higgs and gauge fields, we must couple it to the Bogomol'nyi equations and solve them together as a system. To the best of our knowledge, none of these equations are integrable and we must resort to numerical methods. 

\begin{figure}[htb]
			\centering
			\includegraphics[width=0.45\linewidth]{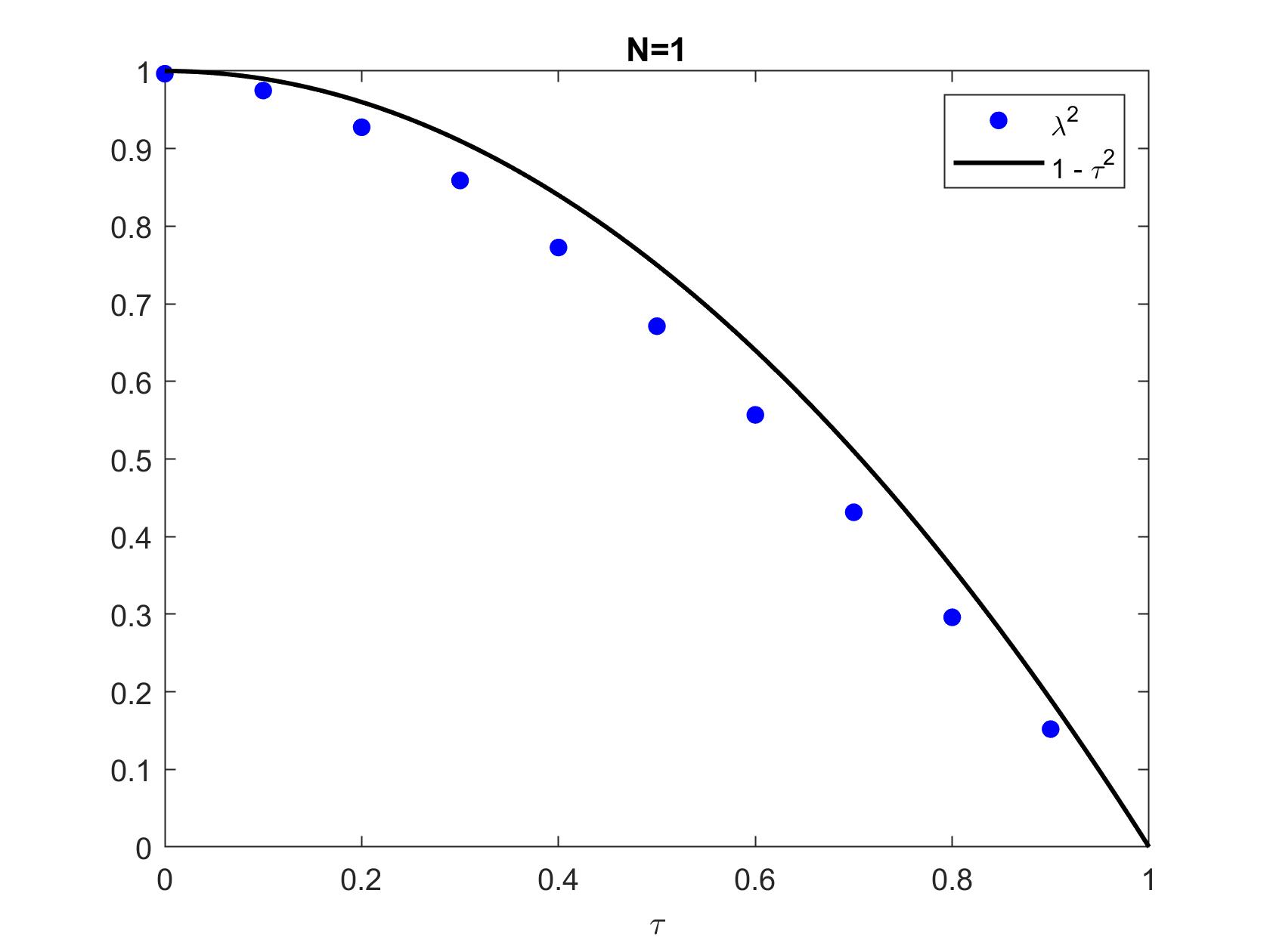}
            \includegraphics[width=0.45\linewidth]{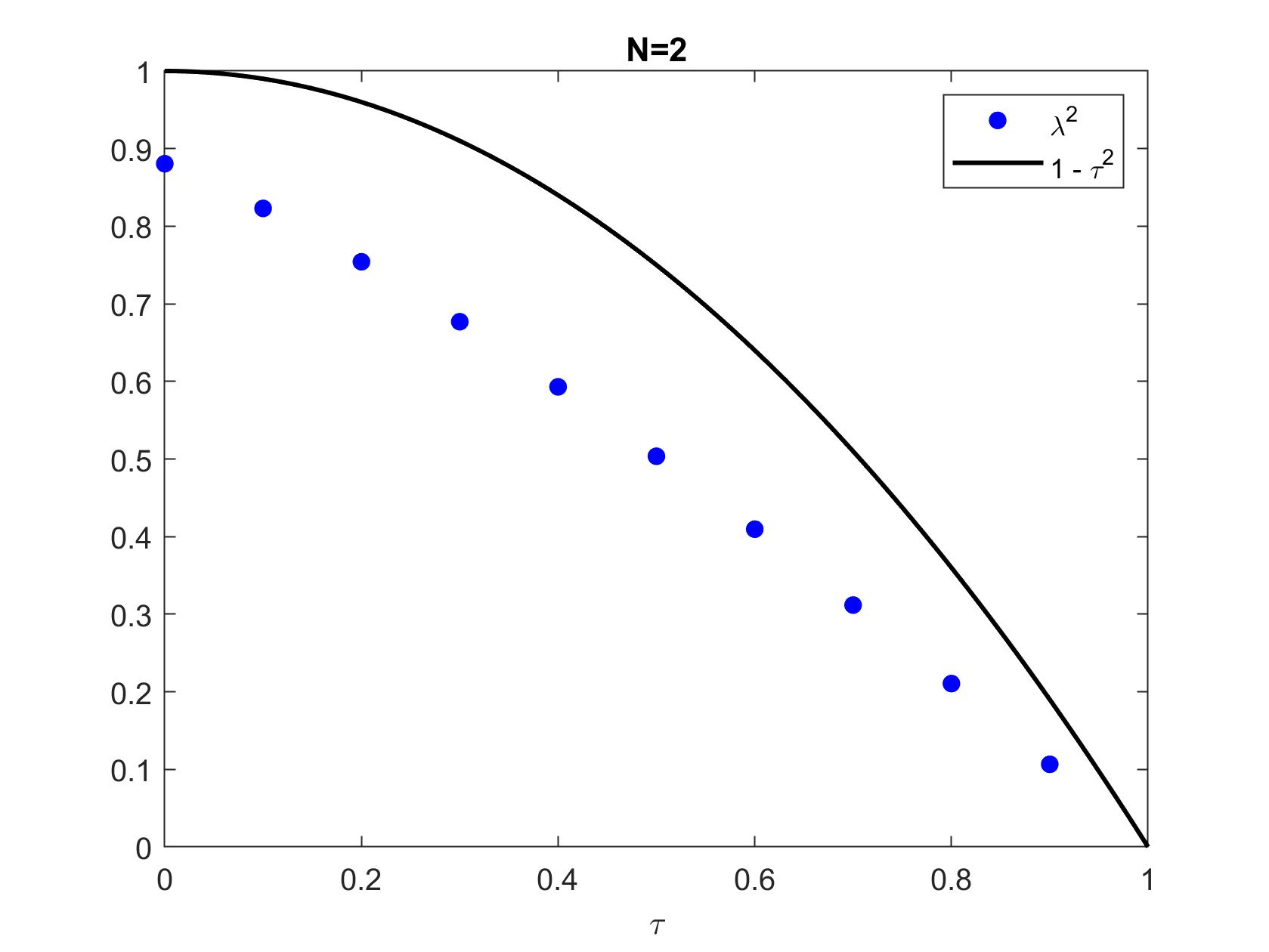}
            \includegraphics[width=0.45\linewidth]{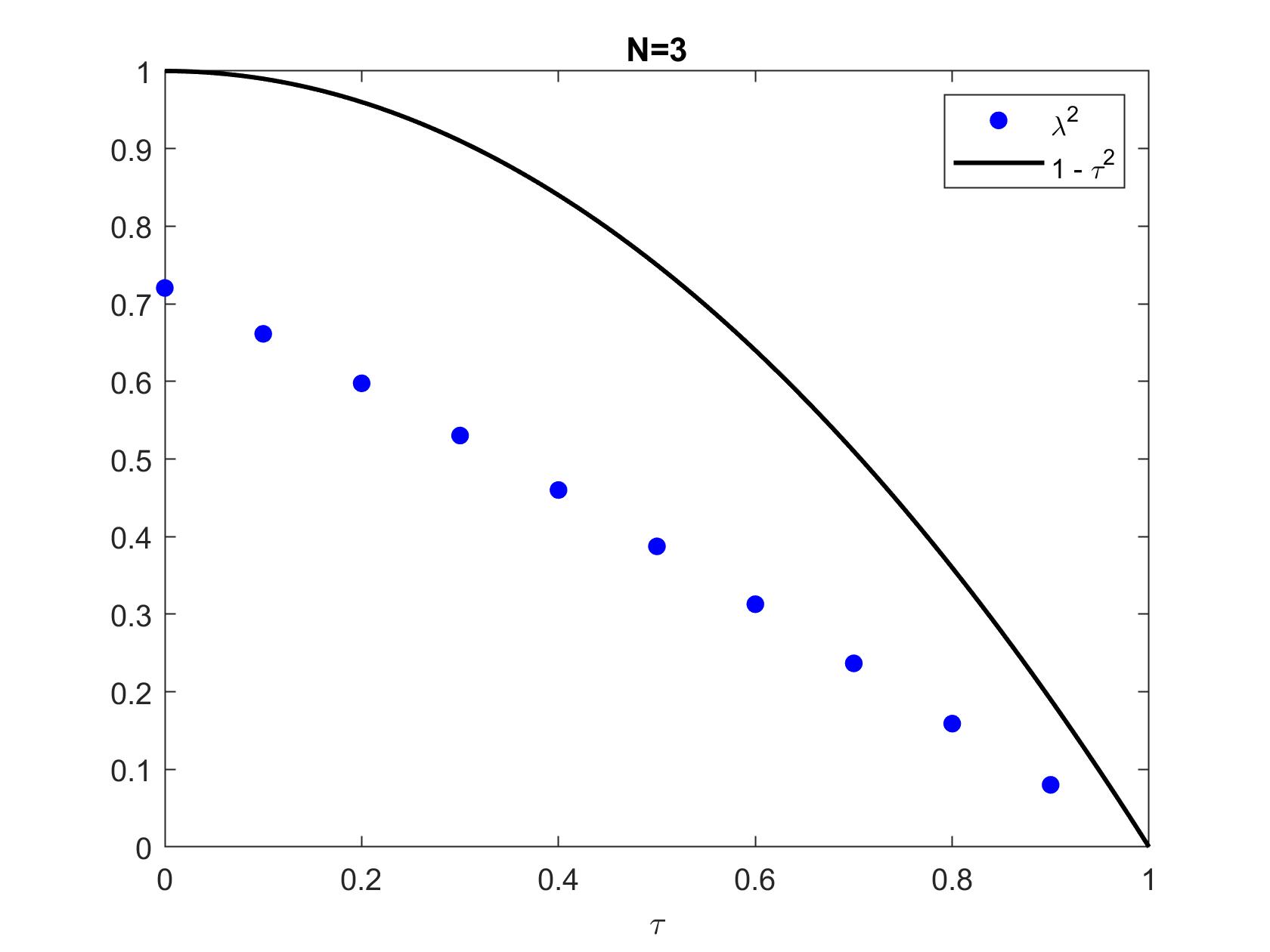}
			\caption{The eigenvalue \(\lambda^2\) of the Jacobi operator vs \(\tau\) for a North vortex solution with \(N=1\), \(N=2\) and \(N=3\). In each case, the system was solved for \(k=0\). The black curve represents the scattering threshold \(1-\tau^2\).}
			\label{lambda_vs_tau}
\end{figure}

To solve the Bogomol'nyi equations coupled to the Schr\(\ddot{\text{o}}\)dinger equation for \(\psi(r)\), we firstly build two solutions, one starting from the left, with initial conditions given by the asymptotics for small \(r\) in (\ref{small_r_asymp_a}), (\ref{small_r_asymp_f}), (\ref{small_r_asymp_p}), and one starting from the right, with initial conditions given by the asymptotics for large \(r\) in (\ref{large_r_asymp_a}), (\ref{large_r_asymp_f}), (\ref{large_r_asymp_p}). We integrate the left solution over the range \([r_0,r_1]\), and the right solution over \([r_1, r_{\text{max}}]\). We usually take \(r_0=10^{-6}\), \(r_1=2\), and \(r_{max}=20\). Recall the asymptotics depend on the constants \(A\), \(q\), \(C\) and \(\lambda\) which can be treated as shooting parameters. To match the left and right solutions at the point \(r=r_1\), we find the shooting constants \(A\), \(q\), \(C\) for which the left and right profiles for \(f(r)\), \(a(r)\) and \(\psi'(r)\) agree within \(10^{-7}\) error using the Newton-Raphson method, and simultaneously apply a bisection method to find the value of \(\lambda\) for which the left and right profiles for \(\psi(r)\) agree within the same error. A separate bisection method is required to find the eigenvalue due to the fact that for \(N=1\) it is always very close to the scattering threshold (see first plot of Figure \ref{lambda_vs_tau}), and therefore incorporating it in the Newton-Raphson numerical scheme can lead to non-convergence. 

The system of ODEs is solved using the MATLAB function ode45, which is based on a Runge-Kutta (4,5) method. 

\begin{figure}[htb]
			\centering
			\includegraphics[width=0.4\linewidth]{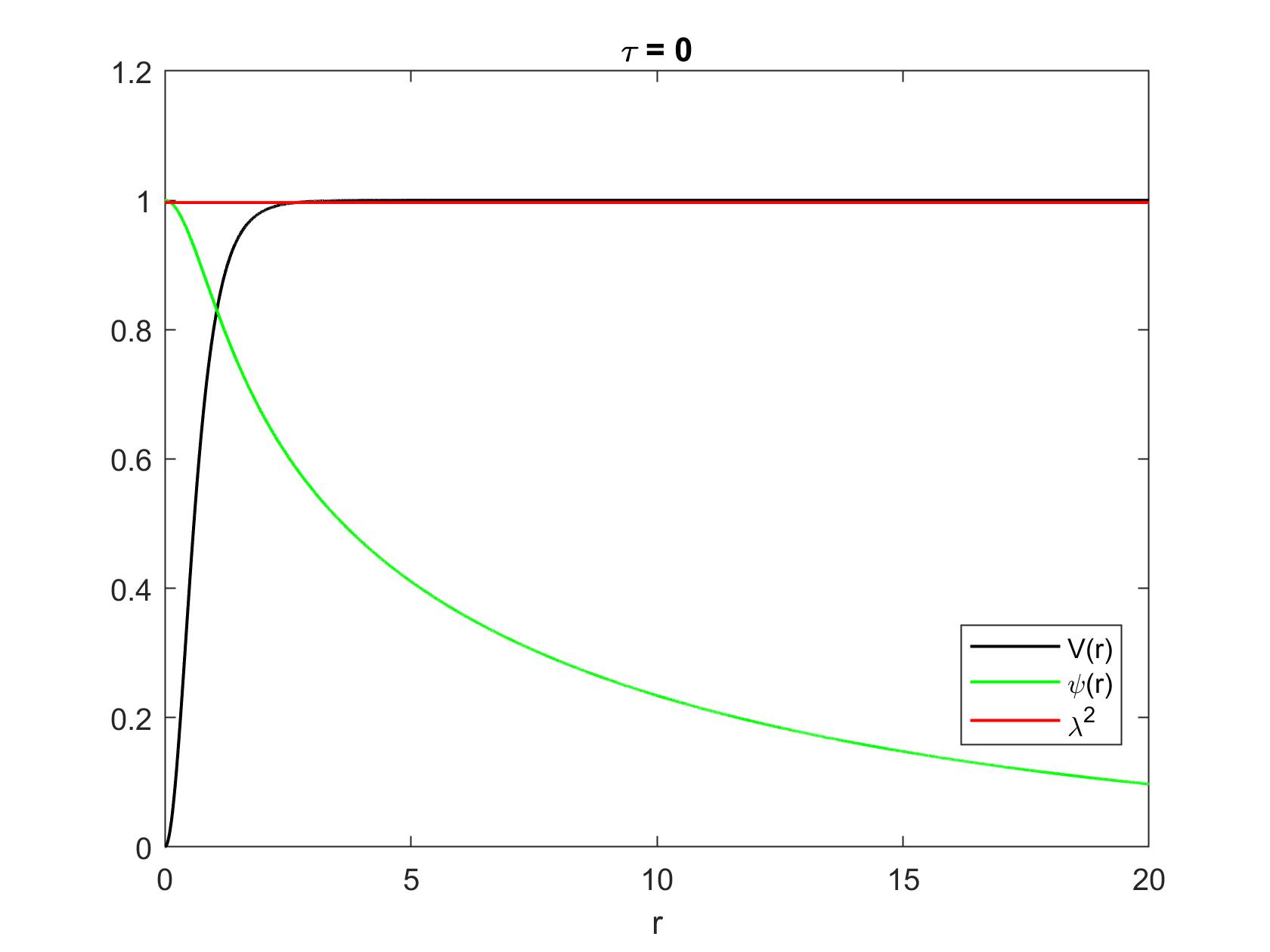}
            \includegraphics[width=0.4\linewidth]{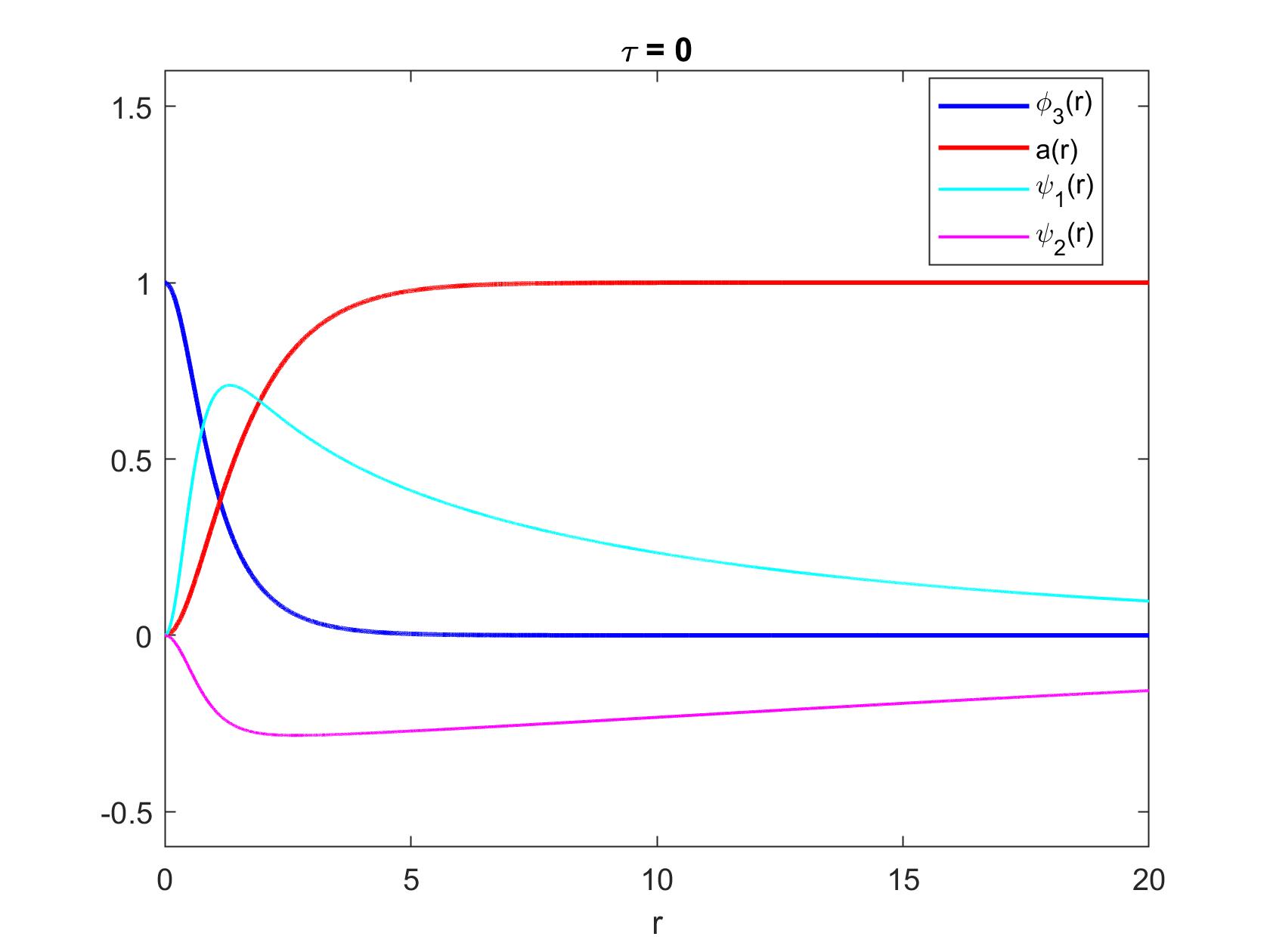}
            \includegraphics[width=0.4\linewidth]{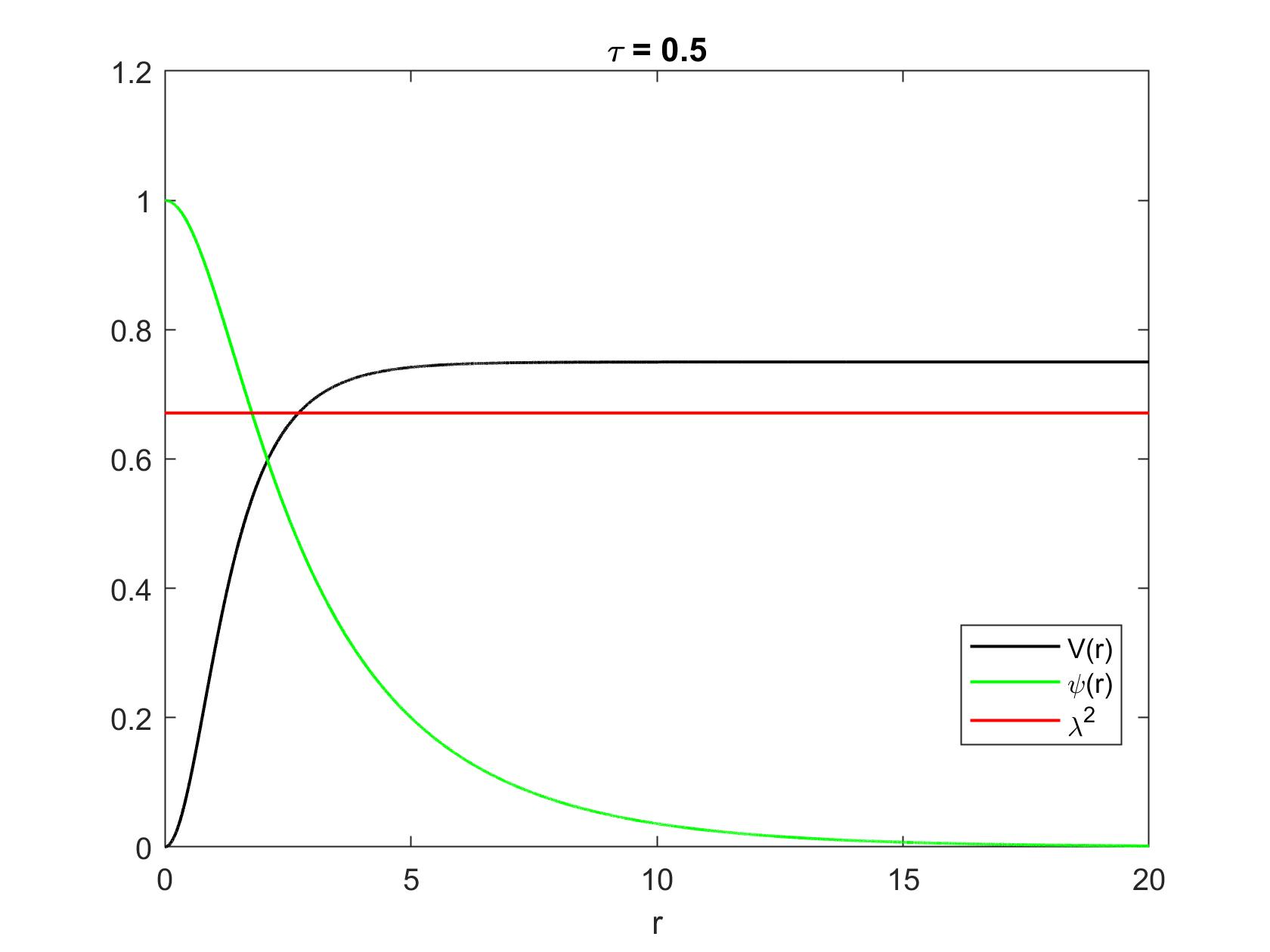}
            \includegraphics[width=0.4\linewidth]{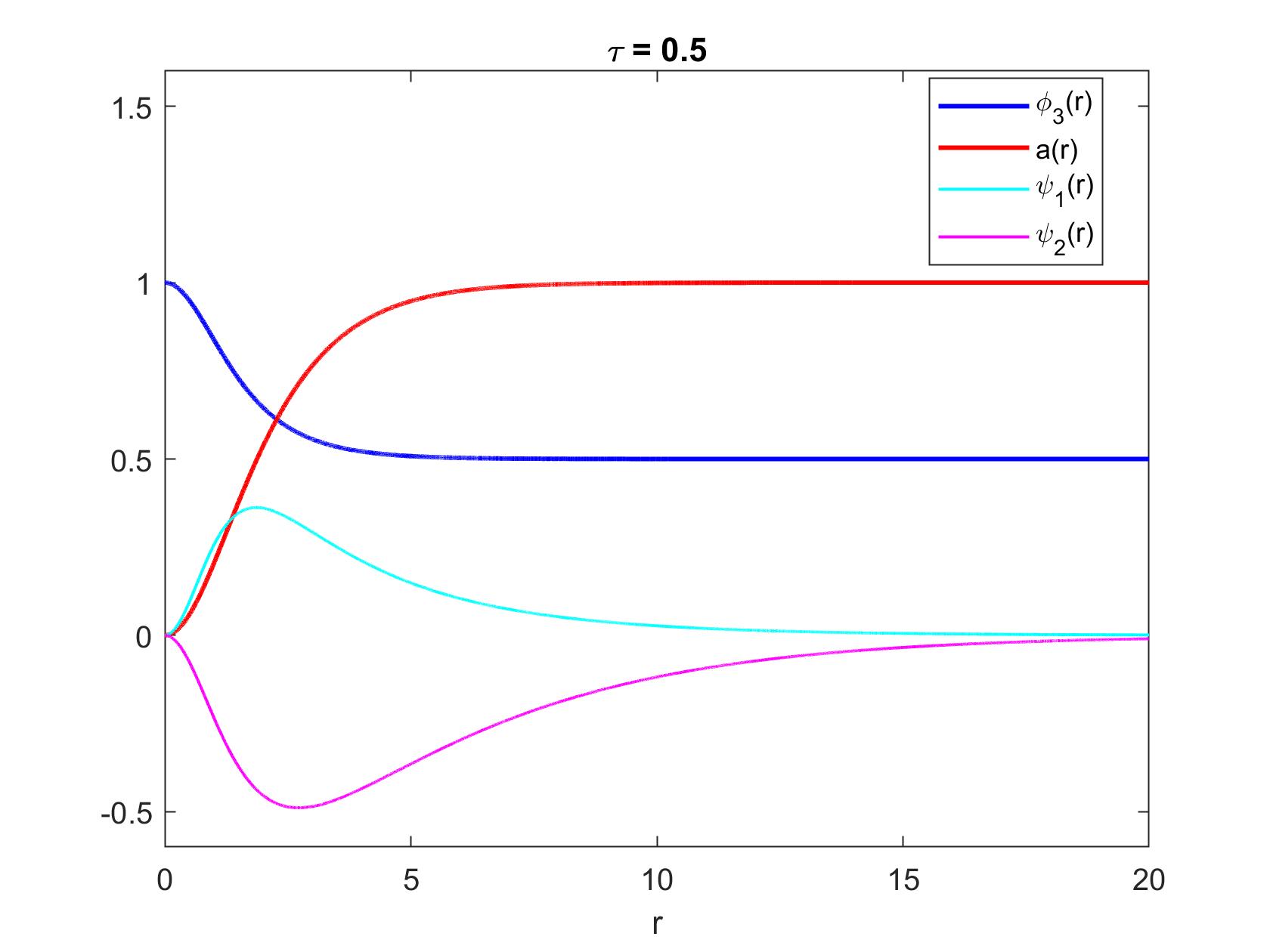}
            \includegraphics[width=0.4\linewidth]{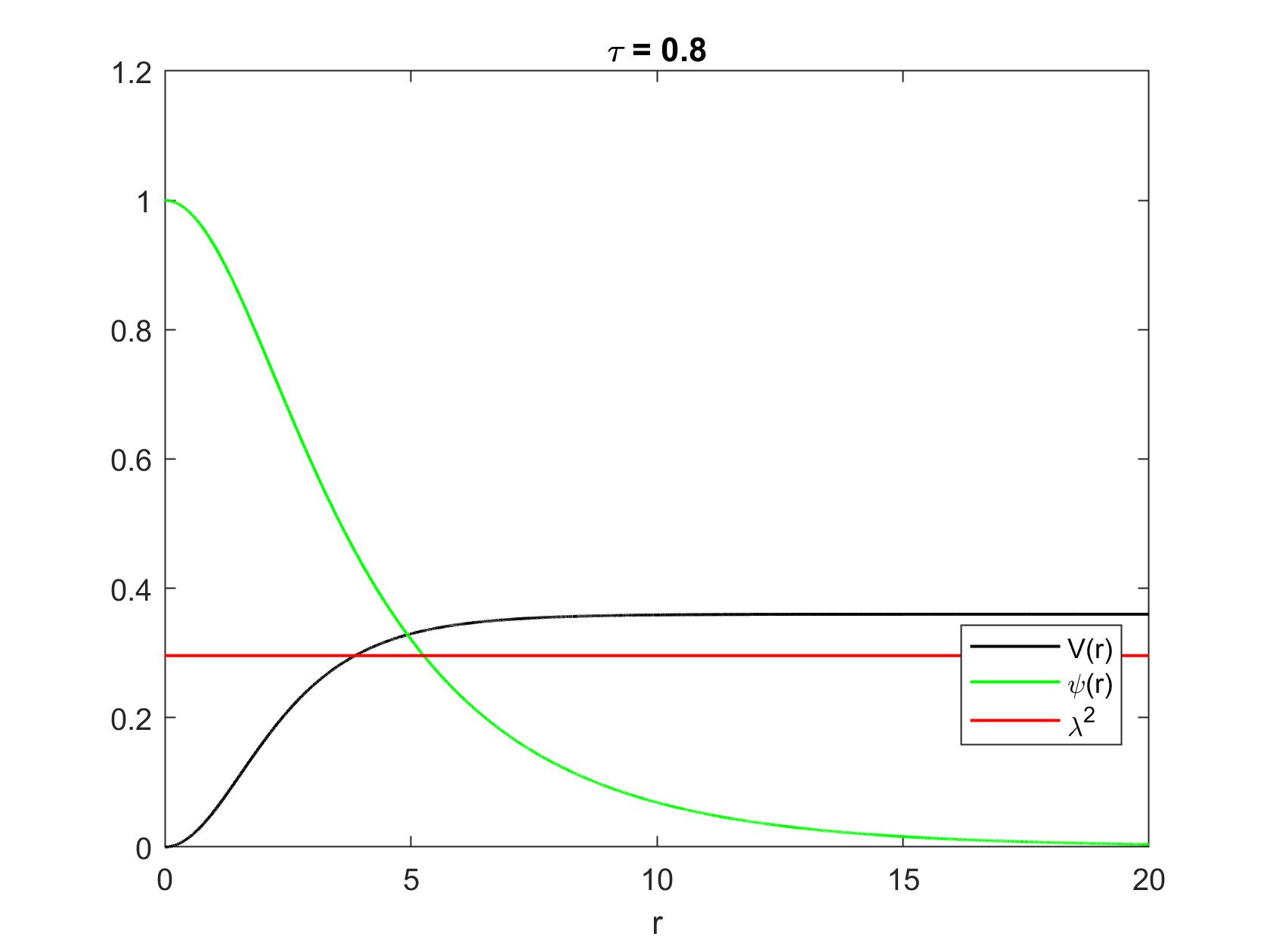}
            \includegraphics[width=0.4\linewidth]{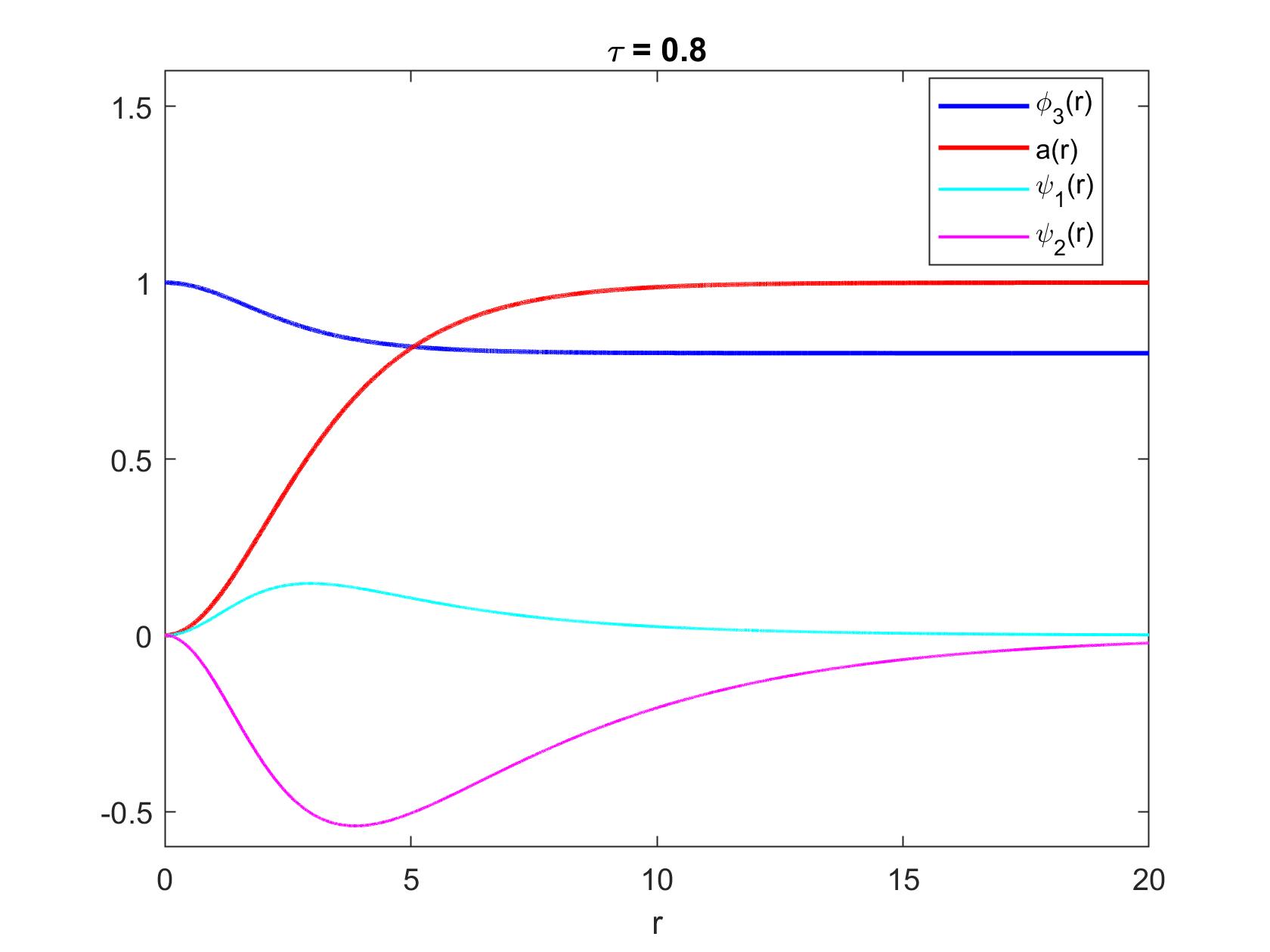}
			\caption{On the left column we present the wavefunction \(\psi(r)\) and the potential of the Schr\(\ddot{\text{o}}\)dinger equation (\ref{radial_schrodinger_p}), along with the eigenvalue \(\lambda^2\). On the right column we present the radial profiles of the gauge field \(a(r)\) and the gauge invariant quantity \(\phi_3=\cos{f(r)}\), along with the shape mode perturbations \(\psi_1(r)\) and \(\psi_2(r)\) given by \(\mathscr{S}_1\mathscr{G} \Psi(r)\), see (\ref{def_psi1}) and (\ref{def_psi2}) for their explicit definition. All quantities were computed for a North vortex solution with \(N=1\), \(k=0\), and different values of \(\tau\), chosen to be \(0\), \(0.5\) and  \(0.8\). The eigenvalues were computed numerically to be \(\lambda^2\approx 0.99654\), \(\lambda^2\approx 0.67116\), \(\lambda^2\approx 0.29596\), respectively.}
			\label{profiles_N=1}
\end{figure}

\begin{figure}[htb]
			\centering
			\includegraphics[width=0.4\linewidth]{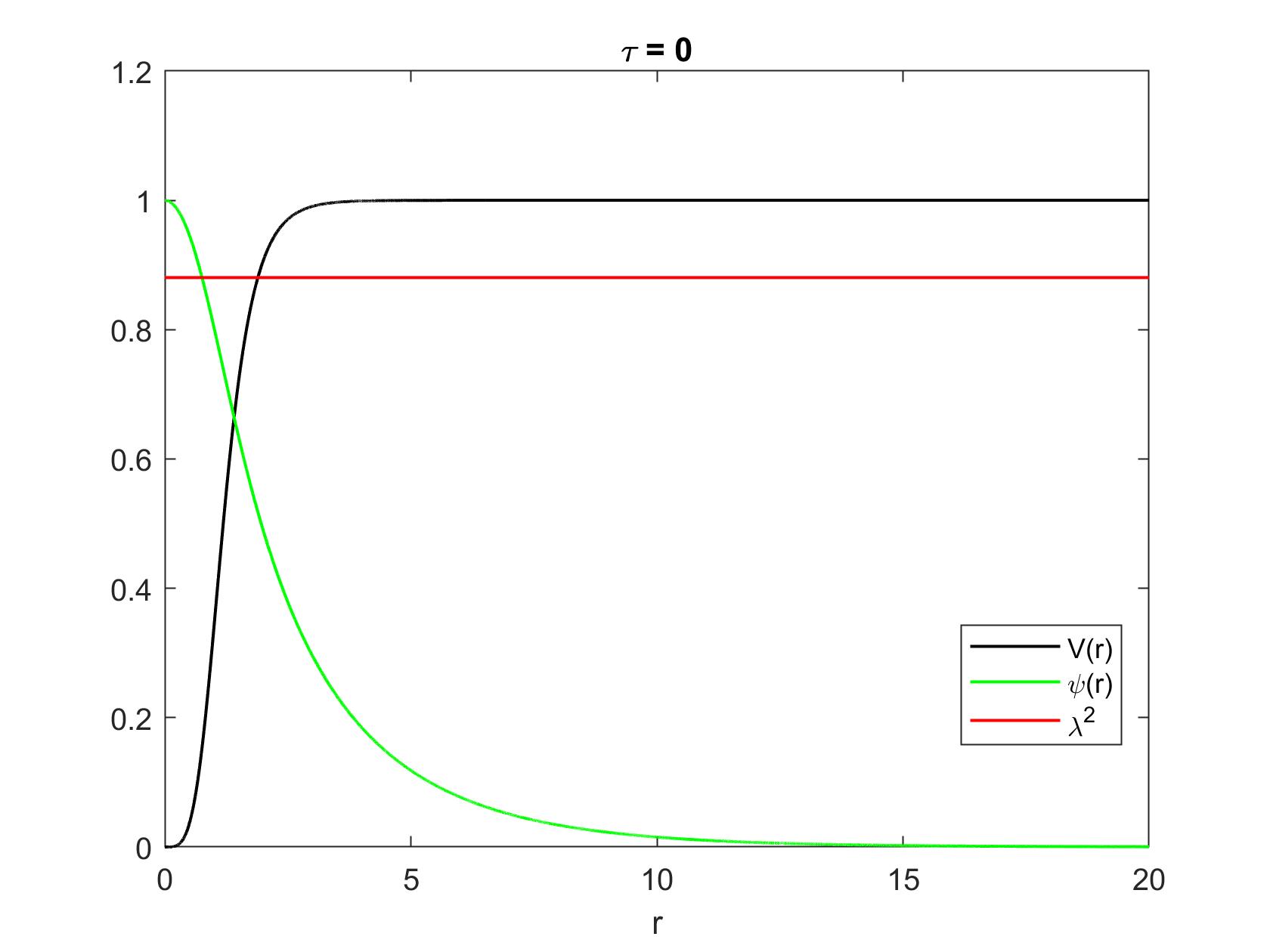}
            \includegraphics[width=0.4\linewidth]{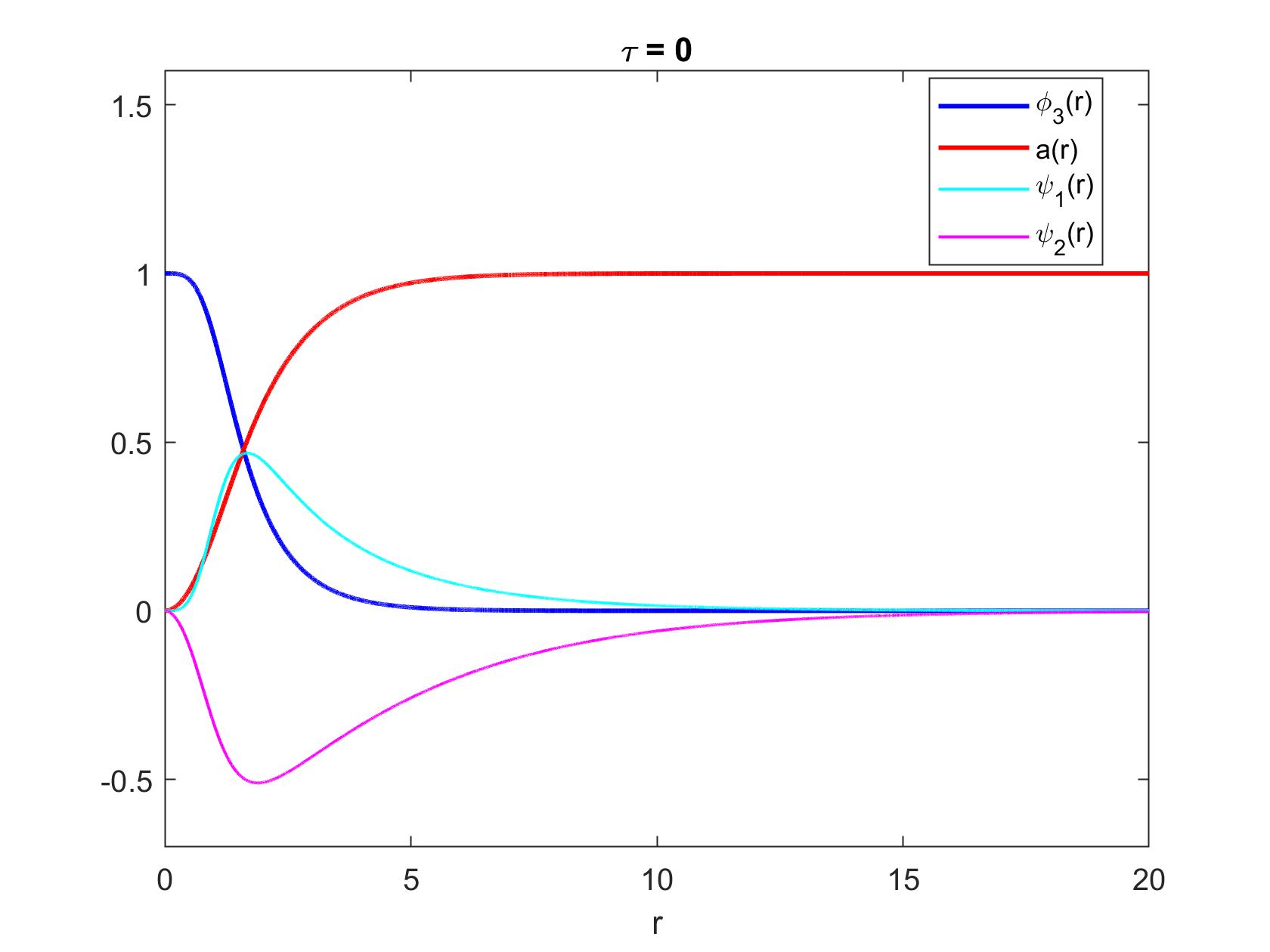}
            \includegraphics[width=0.4\linewidth]{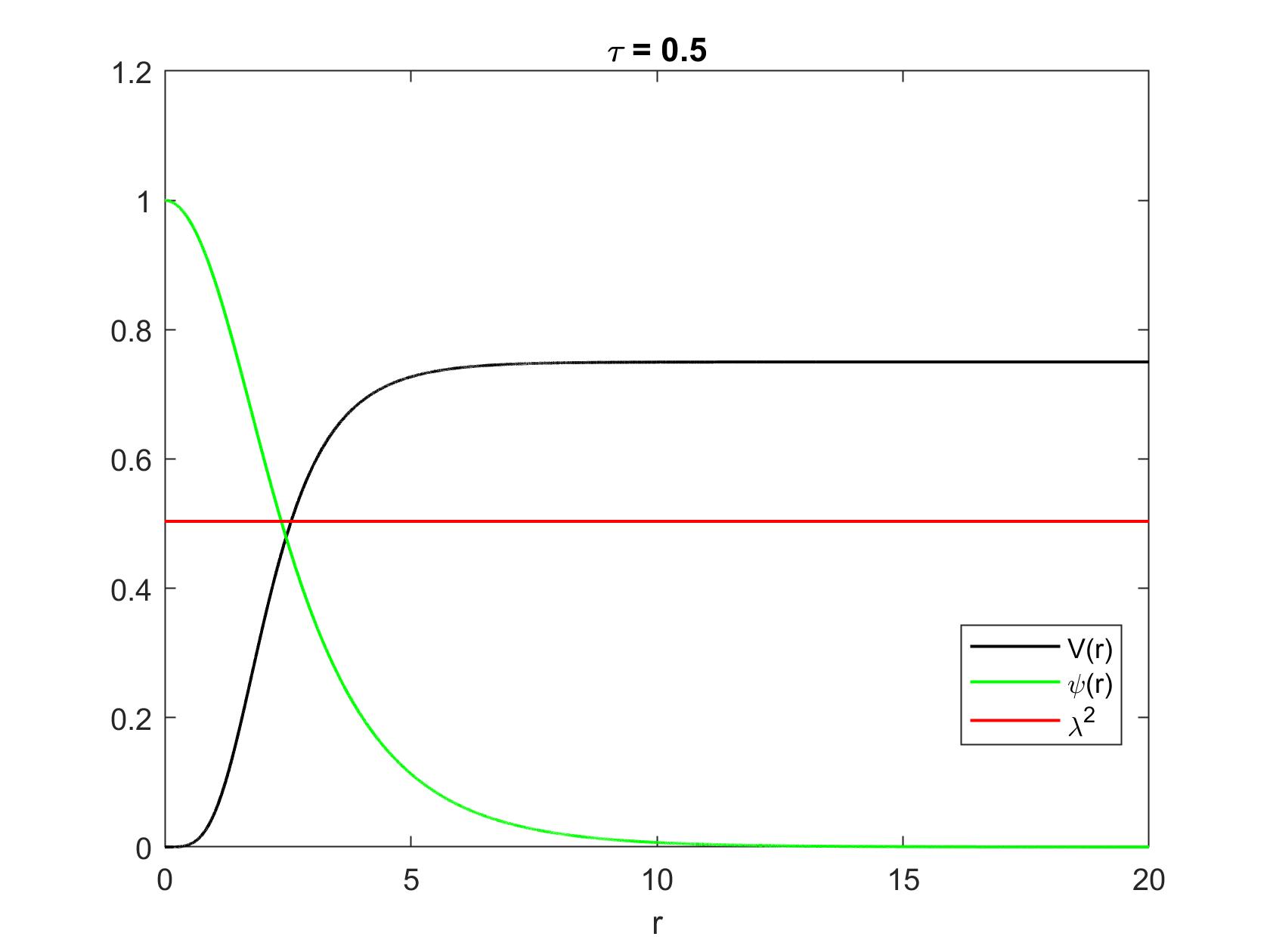}
            \includegraphics[width=0.4\linewidth]{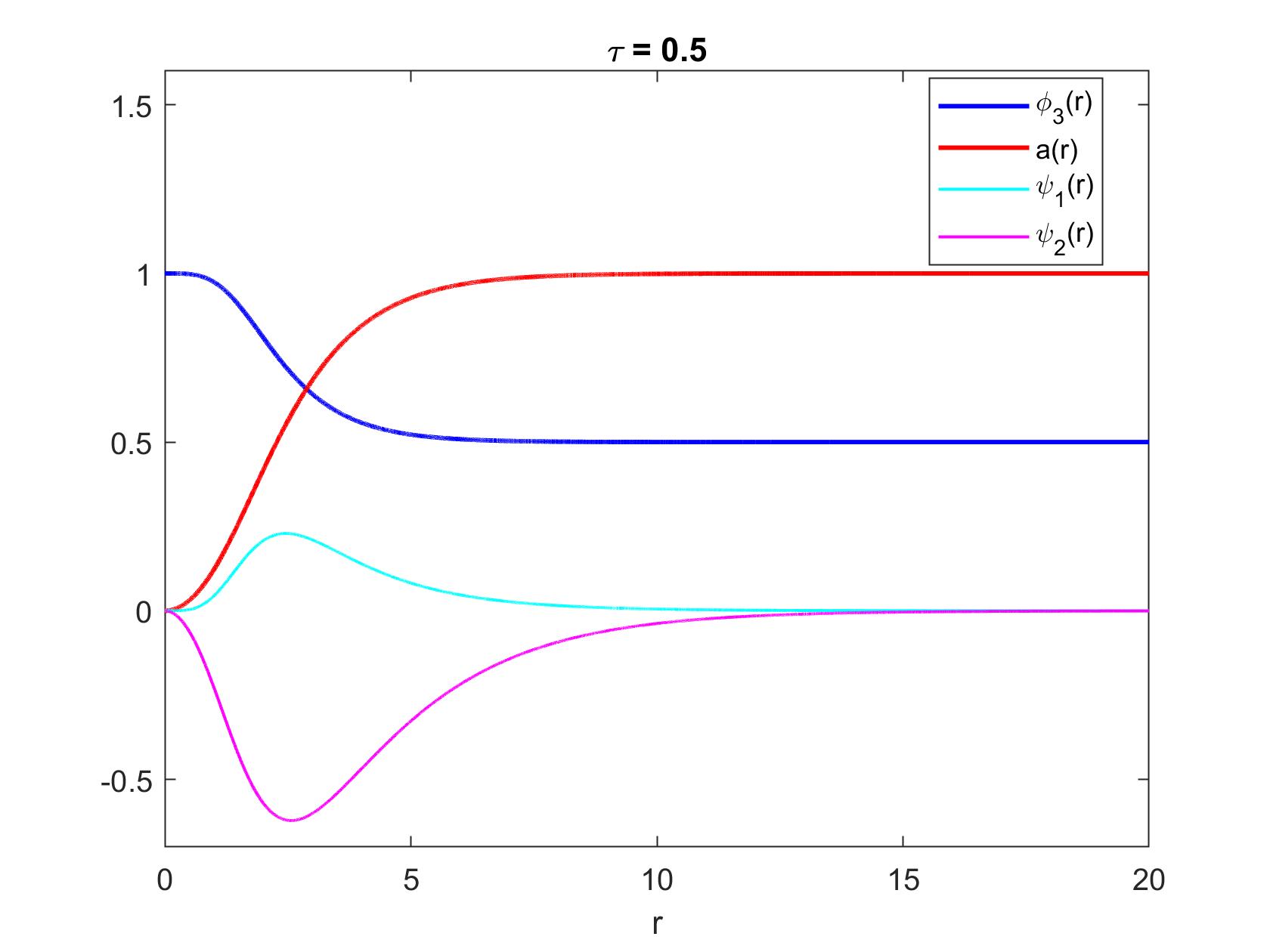}
            \includegraphics[width=0.4\linewidth]{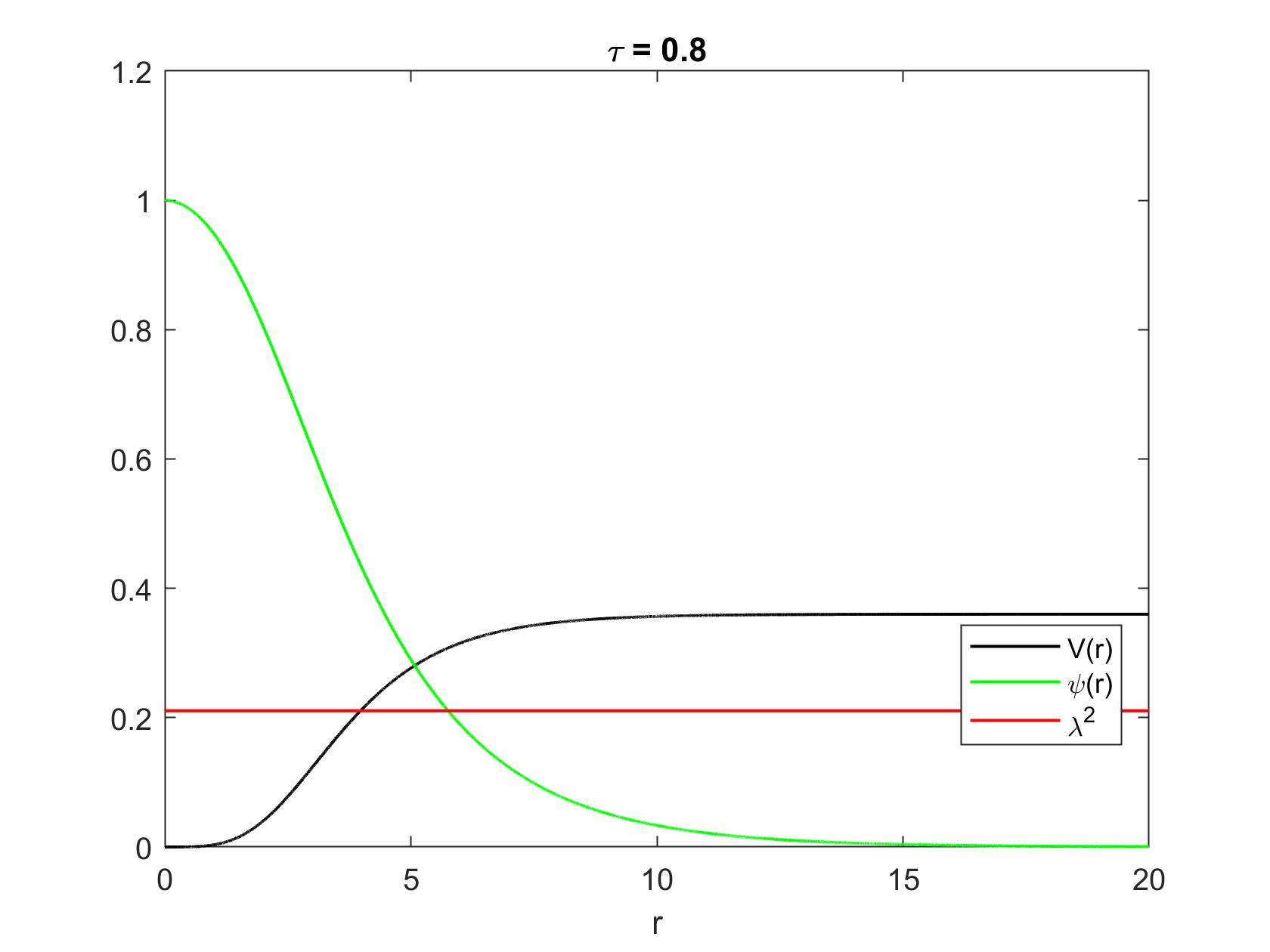}
            \includegraphics[width=0.4\linewidth]{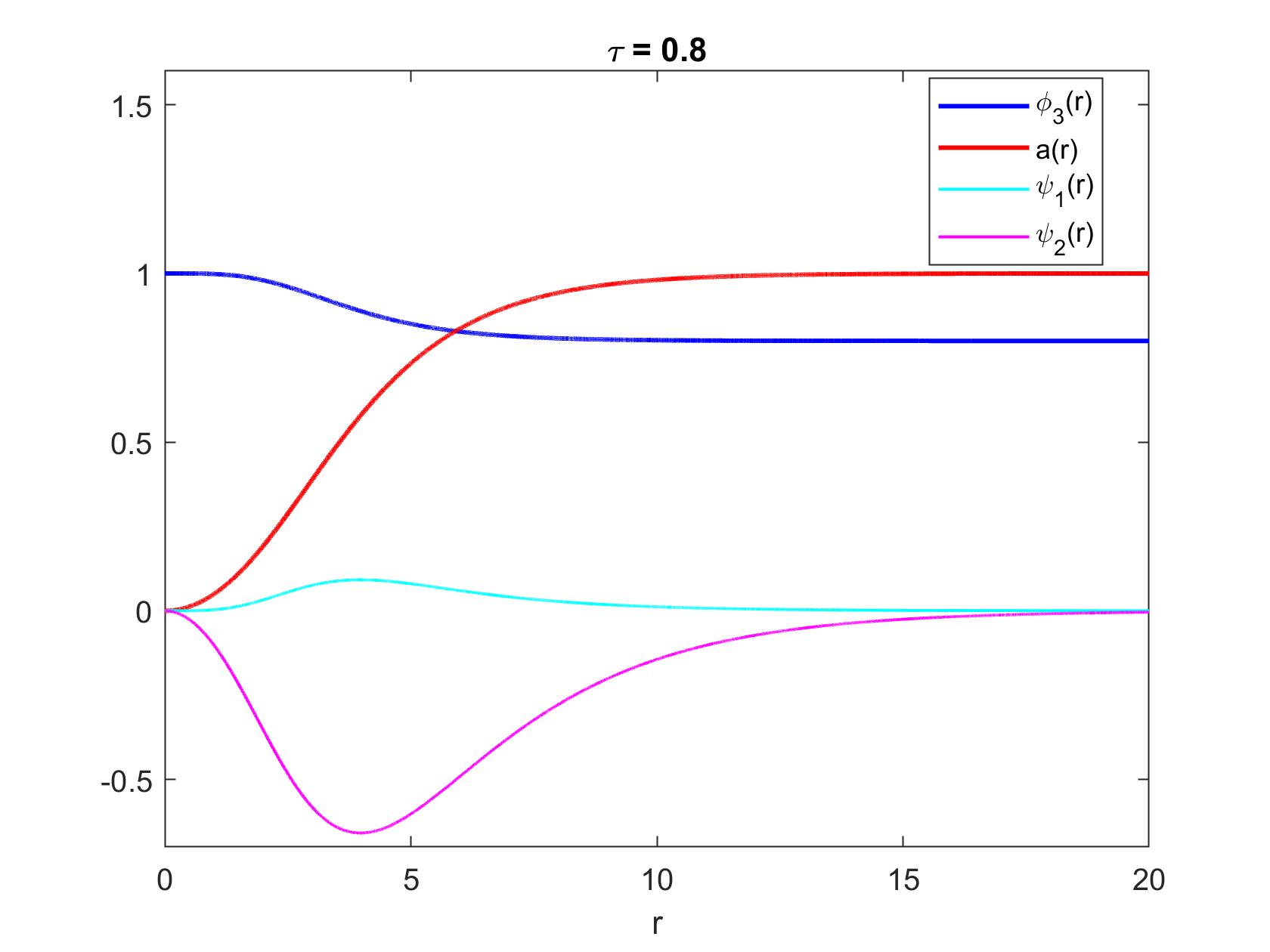}
			\caption{On the left column we present the wavefunction \(\psi(r)\) and the potential of the Schr\(\ddot{\text{o}}\)dinger equation (\ref{radial_schrodinger_p}), along with the eigenvalue \(\lambda^2\). On the right column we present the radial profiles of the gauge field \(a(r)\) and the gauge invariant quantity \(\phi_3=\cos{f(r)}\), along with the shape mode perturbations \(\psi_1(r)\) and \(\psi_2(r)\) given by \(\mathscr{S}_1\mathscr{G} \Psi(r)\), see (\ref{def_psi1}) and (\ref{def_psi2}) for their explicit definition. All quantities were computed for a North vortex solution with \(N=2\), \(k=0\), and different values of \(\tau\), chosen to be \(0\), \(0.5\) and  \(0.8\). The eigenvalues were computed numerically to be \(\lambda^2\approx 0.88054\), \(\lambda^2\approx 0.50355\), \(\lambda^2\approx 0.21058\), respectively.}
			\label{profiles_N=2}
\end{figure}

\begin{figure}[htb]
			\centering
			\includegraphics[width=0.4\linewidth]{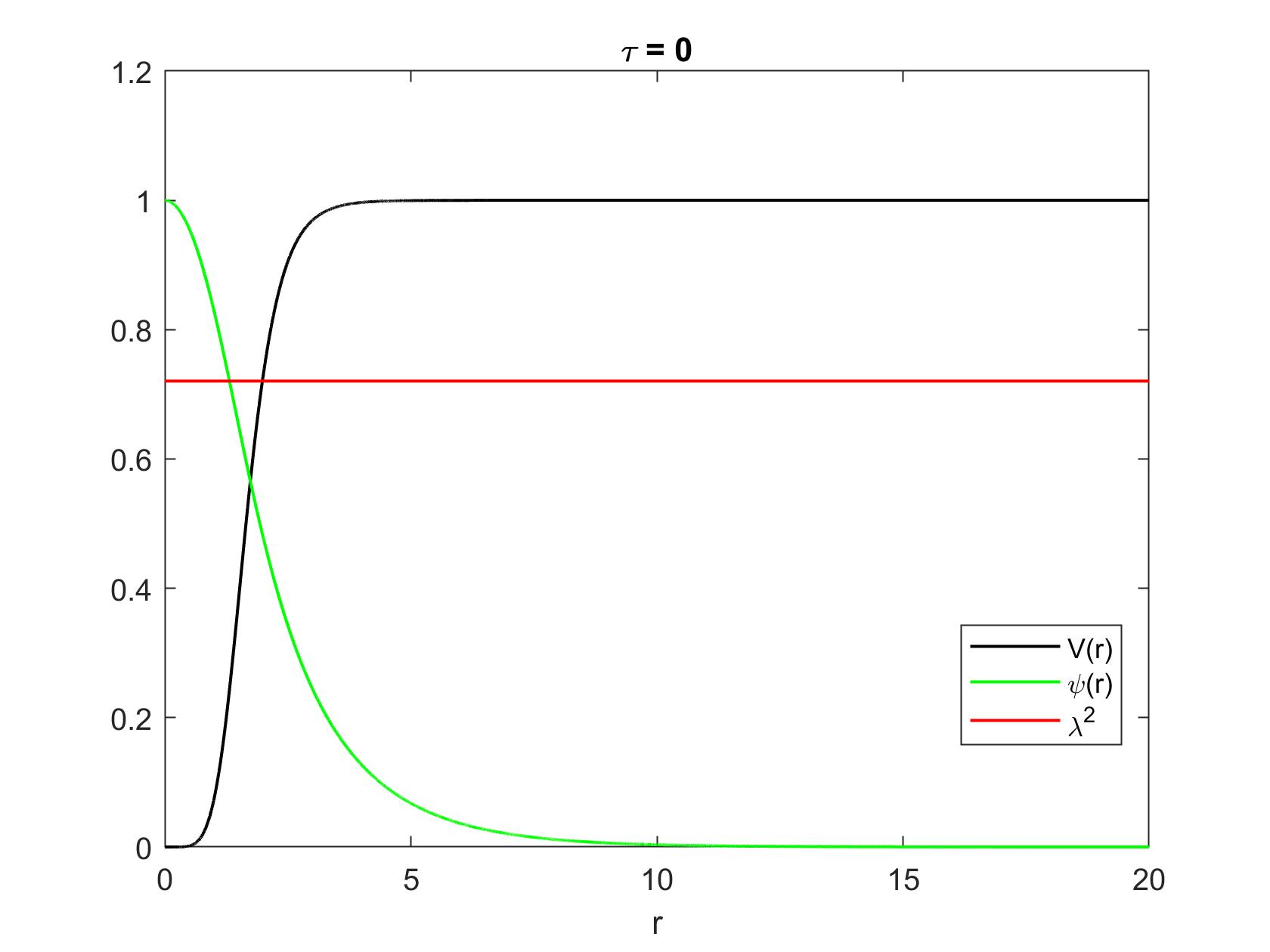}
            \includegraphics[width=0.4\linewidth]{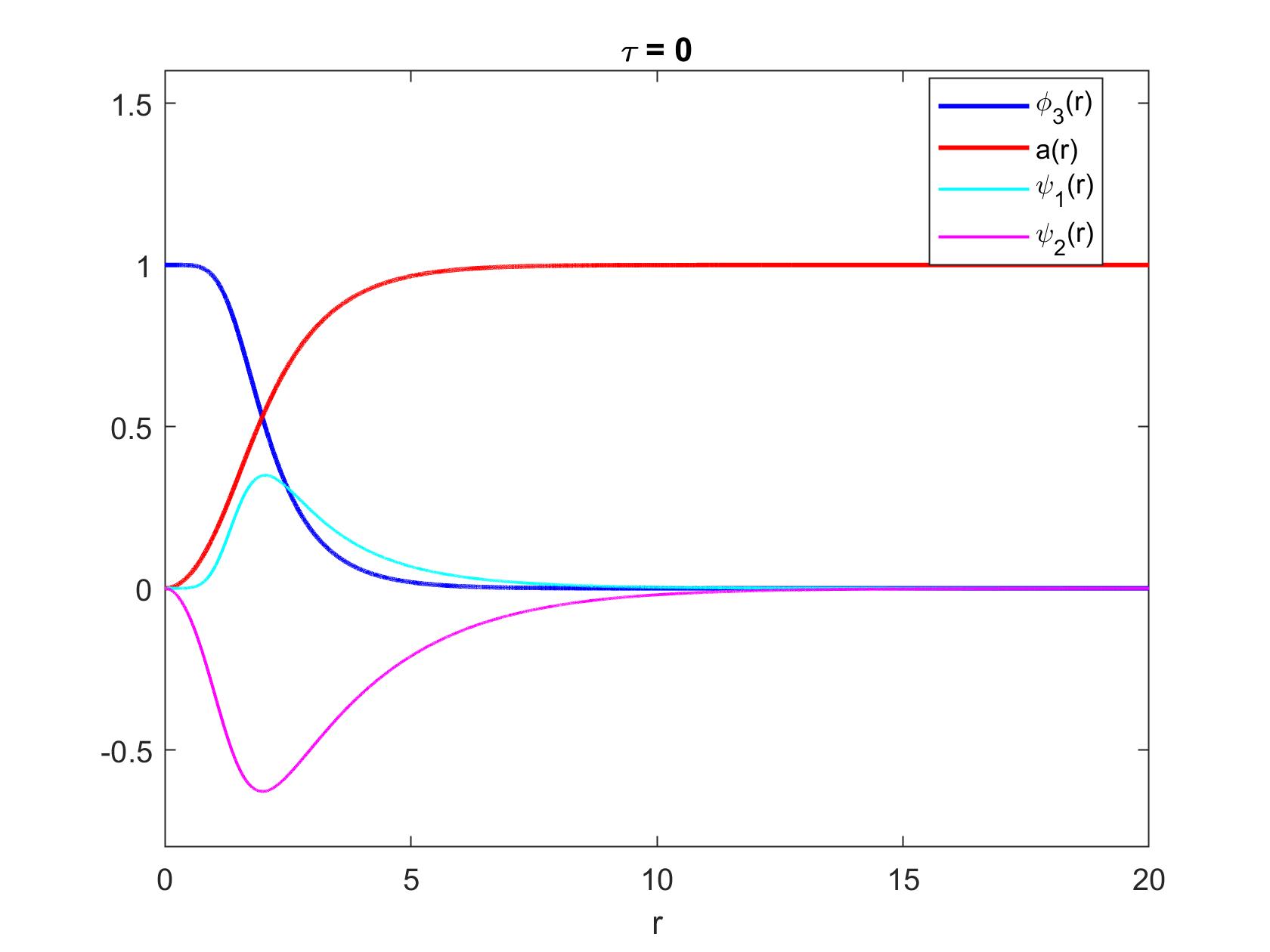}
            \includegraphics[width=0.4\linewidth]{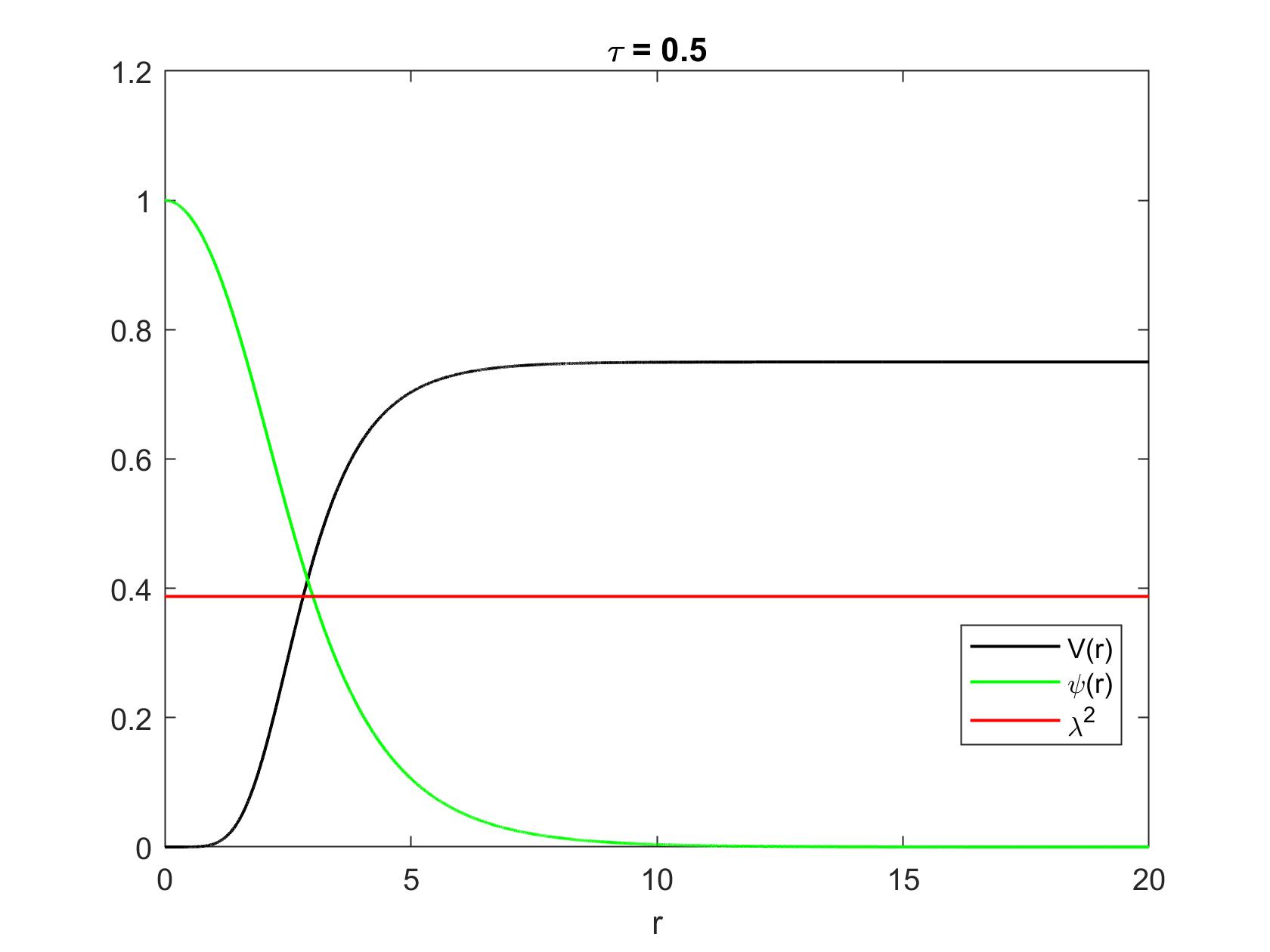}
            \includegraphics[width=0.4\linewidth]{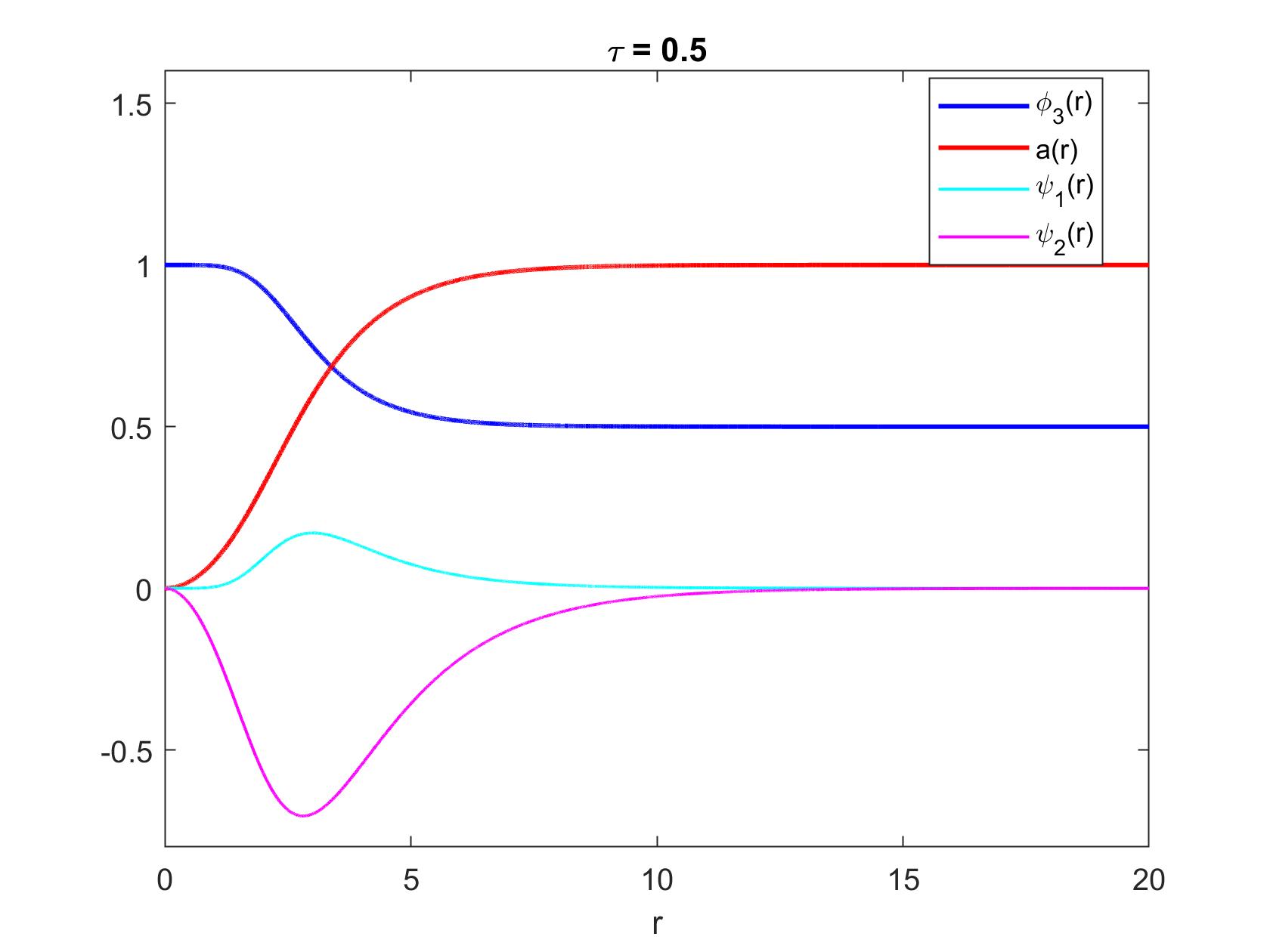}
            \includegraphics[width=0.4\linewidth]{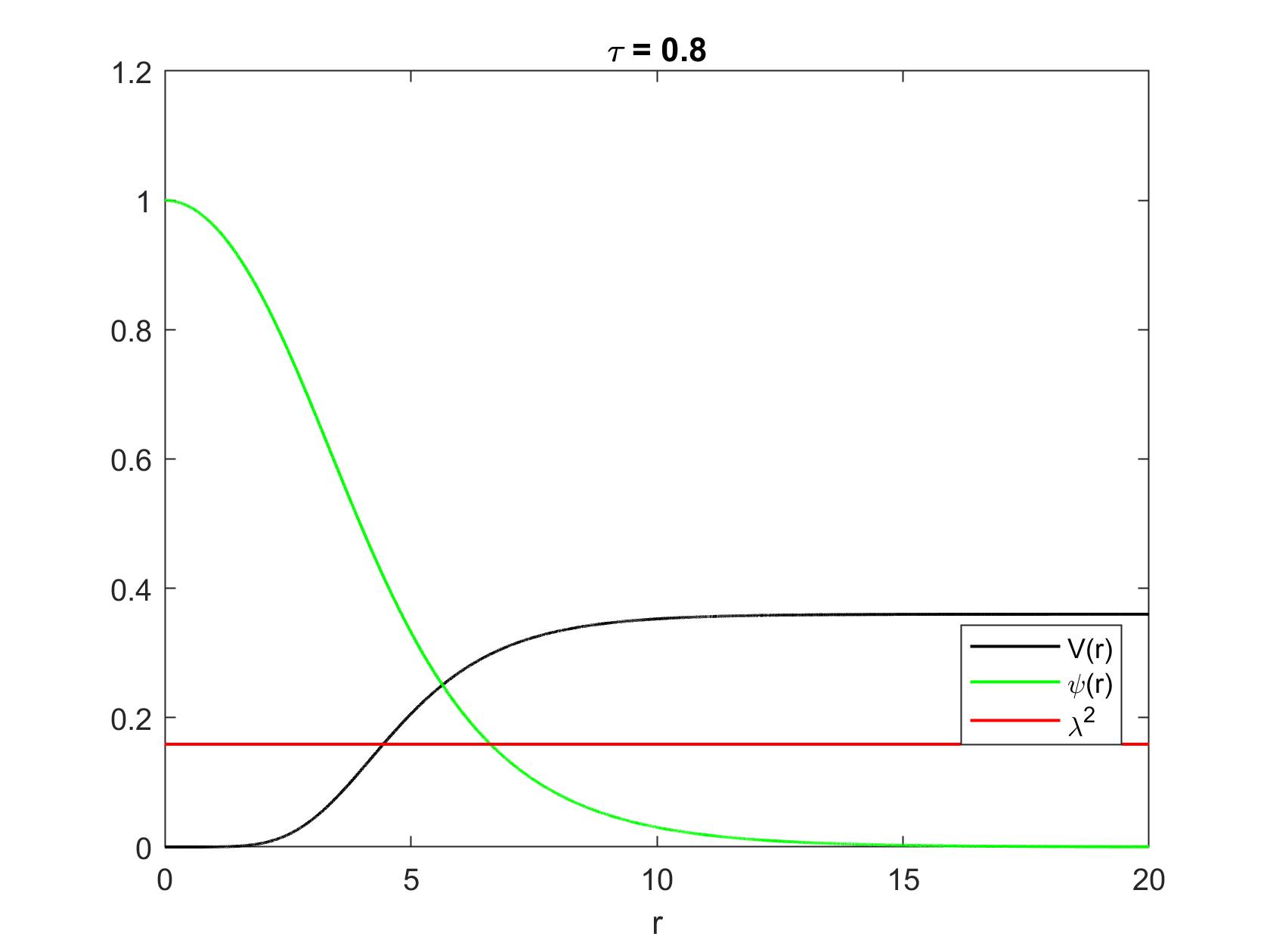}
            \includegraphics[width=0.4\linewidth]{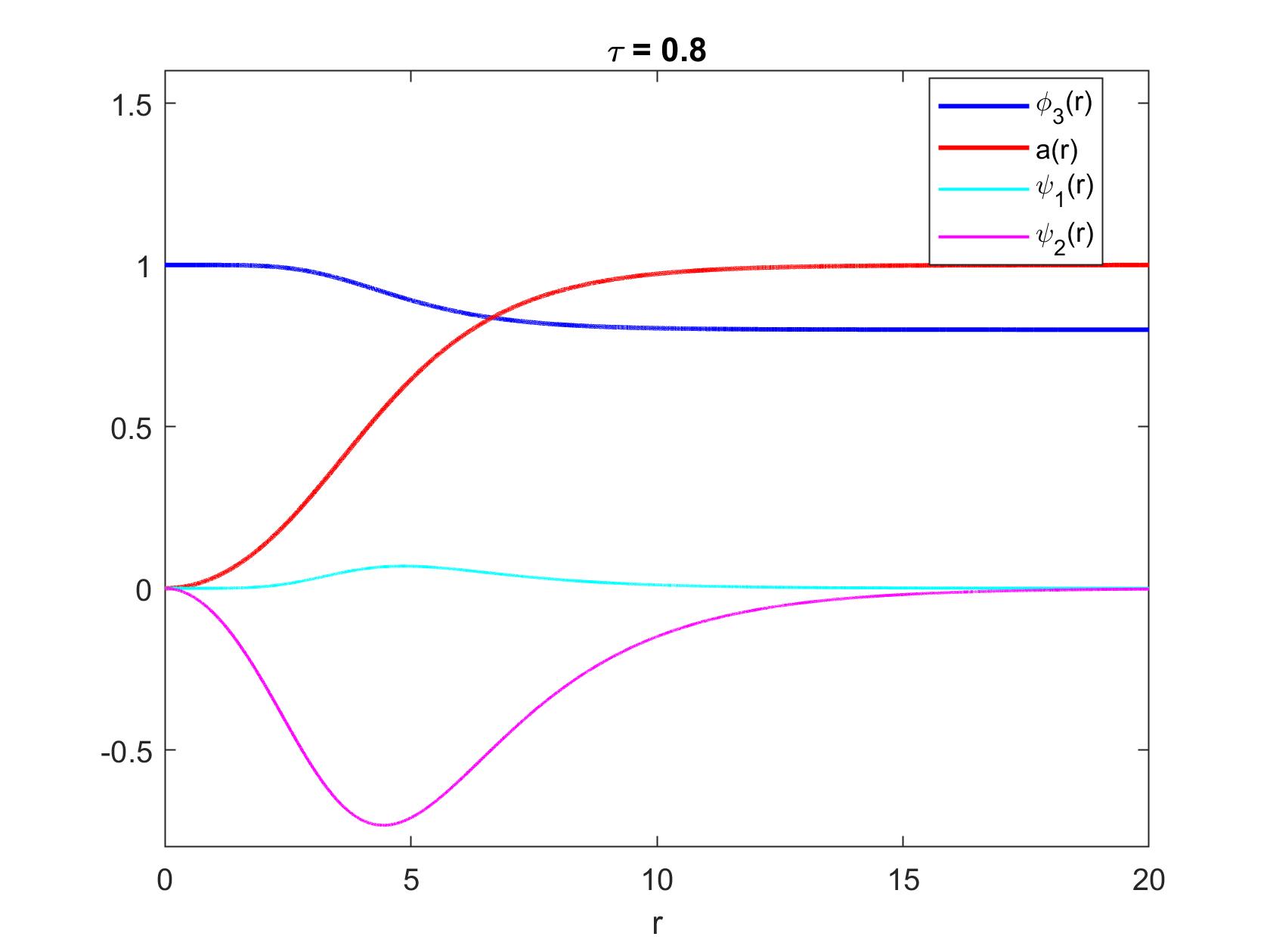}
			\caption{On the left column we present the wavefunction \(\psi(r)\) and the potential of the Schr\(\ddot{\text{o}}\)dinger equation (\ref{radial_schrodinger_p}), along with the eigenvalue \(\lambda^2\). On the right column we present the radial profiles of the gauge field \(a(r)\) and the gauge invariant quantity \(\phi_3=\cos{f(r)}\), along with the shape mode perturbations \(\psi_1(r)\) and \(\psi_2(r)\) given by \(\mathscr{S}_1\mathscr{G} \Psi(r)\), see (\ref{def_psi1}) and (\ref{def_psi2}) for their explicit definition. All quantities were computed for a North vortex solution with \(N=3\), \(k=0\), and different values of \(\tau\), chosen to be \(0\), \(0.5\) and  \(0.8\). The eigenvalues were computed numerically to be \(\lambda^2\approx 0.72042\), \(\lambda^2\approx 0.38743\), \(\lambda^2\approx 0.15896\), respectively.}
			\label{profiles_N=3}
\end{figure}

For the case \(N=1\), the Schr\(\ddot{\text{o}}\)dinger equation (\ref{radial_schrodinger_p}) admits bound states only for \(k=0\). Higher integer values for \(k\) result in potentials which are always above the  bound states threshold \(1-\tau^2\). In Figure \ref{lambda_vs_tau}, we show the eigenvalue \(\lambda^2\) of the Jacobi operator for various values of \(\tau\) in the interval \([0,1)\), for solutions with vortex numbers \(N=1,2,3\). Figures \ref{profiles_N=1}, \ref{profiles_N=2} and \ref{profiles_N=3} show the eigenfunction \(\psi(r)\), the Schr\(\ddot{\text{o}}\)dinger potential, and the shape mode perturbations \(\psi_1(r)\), \(\psi_2(r)\) for the cases \(N=1,2,3\) and specific values of \(\tau\).

\begin{figure}[htb]
			\centering
			\includegraphics[width=0.6\linewidth]{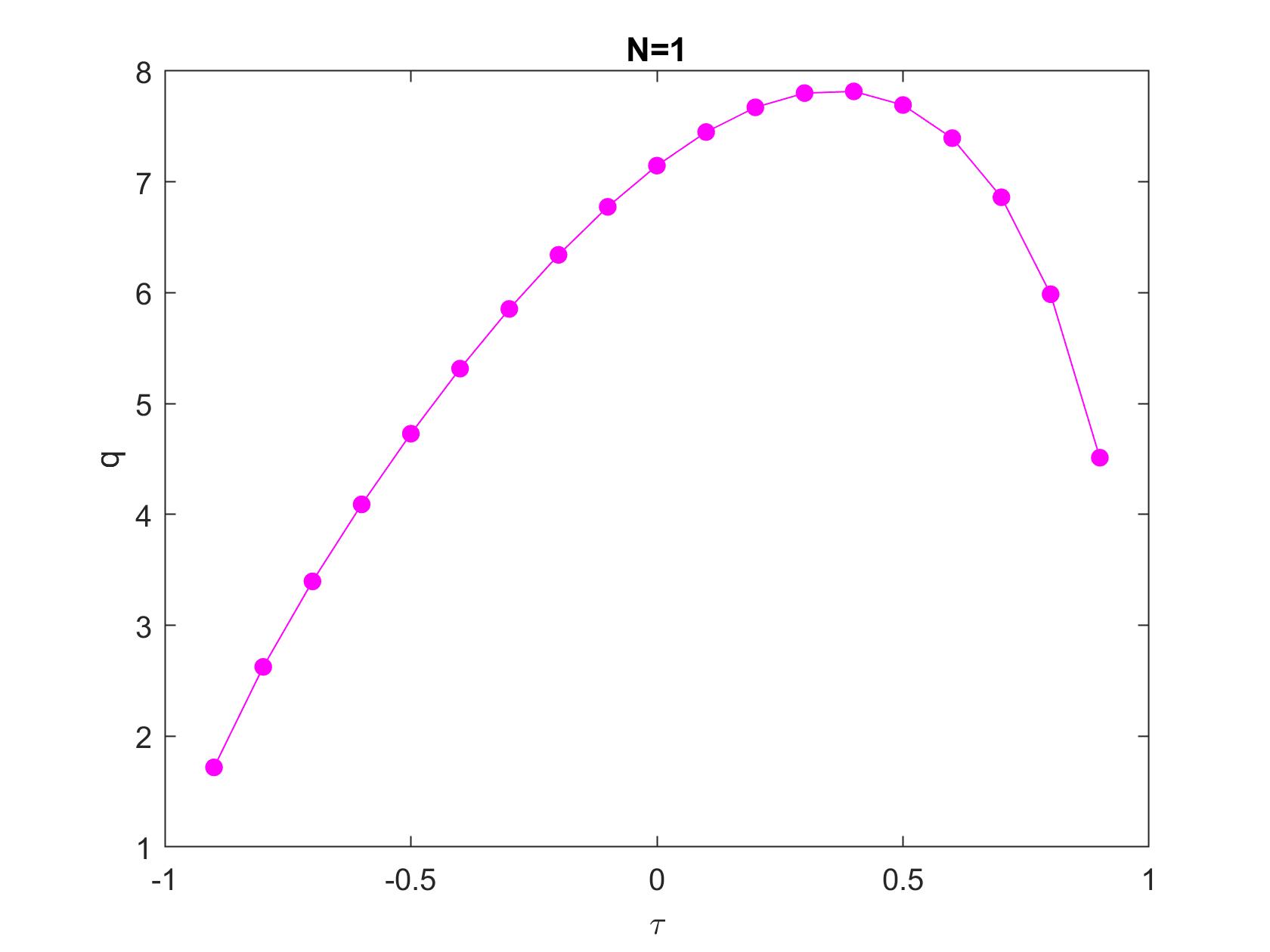}
			\caption{The asymptotic vortex charge \(q\) vs \(\tau\) for the case \(N=1\).}
			\label{q_vs_tau}
\end{figure}

In Theorem \ref{theorem_shapemodes_existence}, we proved that for \(k_-=0\) and \(\tau\in[0,1)\) then there exists at least one shape mode of the vortex solution. This is indeed confirmed by the numerical results presented in Figure \ref{lambda_vs_tau}. In Theorem \ref{theorem_shapemodes_existence_extended}, we further remarked that since the energy depends continuously on \(\tau\), there must be some \(\tau_\star>0\) such that we can extend the range of \(\tau\) for which a shape mode exists to \(\tau\in(-\tau_\star,1)\). Our numerical strategy shows that, in the case of radial vortices with \(N=1\), this threshold is \(\tau_\star\approx 0.001\).

A surprising result is that for a single vortex the eigenvalue \(\lambda^2\) is always very close to the scattering threshold \(1-\tau^2\), see Figure \ref{lambda_vs_tau}. Hence, the binding energy of the shape mode of a 1-vortex for all values of \(\tau\) is very low in magnitude. Even for mid-range values of \(\tau\), for which the gap between the eigenvalue and the threshold appears to increase (see Figures \ref{lambda_vs_tau} and \ref{profiles_N=1}), the binding energy is, in fact, still small. For instance, in the case \(\tau=0.5\), the binding energy is \(E\approx-0.0788\). This does not happen, for example, in the case of the Abelian-Higgs model \cite{dissecting}, where for \(N=1\) the shape mode has eigenvalue \(\approx 0.777476\) and the bound state threshold is 1, giving \(E\approx 0.222524\). This suggests that weakly bound shape modes could be a feature of the \(\mathbb{C}P^1\) model.

In Figure \ref{q_vs_tau} we compute the asymptotic charge \(q\) for a 1-vortex given in \eqref{large_r_asymp_a} and \eqref{large_r_asymp_f} against various values of \(\tau\in[0,1)\). This charge appears in the computation of the conformal factor of the \(L^2-\)metric of the vortex-antivortex moduli space in the large separation regime, as shown in \cite{CP1geometry} for the case \(\tau=0\).

\subsection{Are there any other shape modes?}
An important question is whether the eigenvalue problem (\ref{radial_schrodinger}) gives all the possible shape modes for a radially symmetric \(N-\)vortex. To answer this, we 
note that, if $\xivec$ is an eigensection of $\J$ with eigenvalue $\lambda^2>0$, it is an eigensection of $\J^G=\B^{G\dagger}\B^G$ with the same eigenvalue, and $\B^G\xivec$ (which is not $0$) is an eigensection of $\B^G\B^{G\dagger}$ with the same eigenvalue. Hence it suffices to compute the spectrum of the operator $\B^G\B^{G\dagger}$ acting on
$\B^G(\mathcal{PERT})\subset\mathcal{BOG}$.  

We first note that if $\tilde\xivec=\B^G(\xivec)$, then 
\[\tilde{\boldsymbol{\xi}}=\begin{pmatrix}
\dfrac{1}{\sqrt{2}}\tilde{h}(\boldsymbol{e_\theta}\otimes dr-\boldsymbol{e_r}\otimes rd\theta)+\dfrac{1}{\sqrt{2}}\tilde{g}(\boldsymbol{e_r}\otimes dr+\boldsymbol{e_\theta}\otimes rd\theta)\\
\tilde{\beta} \\
\tilde{\gamma}
\end{pmatrix},\]
where \(dr\), \(rd\theta\) is an orthonormal basis of \(T^{\star}\Sigma\), 
\[\boldsymbol{e_r}=(\cos{f}\cos{N\theta},\cos{f}\sin{N\theta},-\sin{f})\]
\[\boldsymbol{e_\theta}=(\sin{N\theta},-\cos{N\theta},0)\]
is an orthonormal basis of \(\boldsymbol{\phi}^*TS^2\), and \(\tilde{h}\), \(\tilde{g}\), \(\tilde{\beta}\), \(\tilde{\gamma}\) are functions of \(r,\theta\).
Then
\[\mathscr{B}^{G\dagger}\tilde{\boldsymbol{\xi}}=\begin{pmatrix}
-\left(\partial_r \tilde{h}+\dfrac{1}{r}\partial_\theta\tilde{g}+\dfrac{1}{r}\left(1+N(1-a)\cos{f}\right)\tilde{h}+\tilde{\gamma} \sin{f}\right)\boldsymbol{e_\theta}-\\
-\left(\partial_r \tilde{g}-\dfrac{1}{r}\partial_\theta\tilde{h}+\dfrac{1}{r}\left(1+N(1-a)\cos{f}\right)\tilde{g}-\tilde{\beta} \sin{f}\right)\boldsymbol{e_r}\\
\left(\partial_r \tilde{\gamma}+\dfrac{1}{r}\partial_\theta\tilde{\beta}+\tilde{h}\sin{f} \right)dr+
\left(-\partial_r \tilde{\beta}+\dfrac{1}{r}\partial_\theta\tilde{\gamma}+\tilde{g}\sin{f}\right)rd\theta
\end{pmatrix}.\]
Finaly we can compute 
\(
\mathscr{B}^G\mathscr{B}^{G\dagger}\tilde{\boldsymbol{\xi}}=\lambda^2\tilde{\boldsymbol{\xi}}\) to obtain a system of decoupled PDEs:
\[-\nabla^2h+\dfrac{2}{r^2}\left(1+N(1-a)\cos{f}\right)\partial_{\theta}g+\left(\dfrac{1}{r^2}\left(1+N(1-a)\cos{f}\right)^2+1-\tau\cos{f}\right.\]\[\left.+\dfrac{N^2}{r^2}(1-a)^2\sin^2{f}\right)h=\lambda^2h\]
\[-\nabla^2g-\dfrac{2}{r^2}\left(1+N(1-a)\cos{f}\right)\partial_{\theta}h+\left(\dfrac{1}{r^2}\left(1+N(1-a)\cos{f}\right)^2+1-\tau\cos{f}\right.\]\[\left.+\dfrac{N^2}{r^2}(1-a)^2\sin^2{f}\right)g=\lambda^2g\]
\[-\nabla^2\beta+\beta\sin^2{f}=\lambda^2\beta\]
\[-\nabla^2\gamma+\gamma\sin^2{f}=\lambda^2\gamma,\]
where we have dropped the tilde symbol to avoid cumbersome notation.

Since the functions \(h,g,\gamma,\beta\) are periodic in \(\theta\), we can decompose them into Fourier modes
\[h(r,\theta)=\Theta_{Nk}(r)\cos{k\theta}, \text{ } g(r,\theta)=\Theta_{Nk}(r)\sin{k\theta}\]
\[\beta(r,\theta)=\psi_{Nk}(r)\cos{k\theta}, \text{ } \gamma(r,\theta)=\psi_{Nk}(r)\sin{k\theta},\]
where \(k\) is an integer, labeling each Fourier sector. Notice the other half of the Fourier space can be obtained by applying the symmetry $\S_2$ explained in Lemma \ref{jacobi_sym}. The PDEs now reduce to a decoupled pair of second order ODEs for each $k$:
\begin{equation} \label{radial_schrodinger_t}
-\Theta''-\dfrac{1}{r}\Theta'+\left(\dfrac{1}{r^2}\left(k+1+N(1-a)\cos{f}\right)^2+1-\tau\cos{f}+\dfrac{N^2}{r^2}(1-a)^2\sin^2{f}\right)\Theta=\lambda^2\Theta
\end{equation}
\begin{equation} \label{radial_schrodinger_p}
-\psi''-\dfrac{1}{r}\psi'+\left(\dfrac{k^2}{r^2}+\sin^2{f}\right)\psi=\lambda^2\psi,
\end{equation}
where we have dropped the subscripts for \(\psi\) and \(\Theta\). These are both 2D radial Schr\(\ddot{\text{o}}\)dinger equations in a central potential depending on the radial profiles of the Higgs and gauge fields, \(f(r)\) and \(a(r)\), respectively.

We immediately notice that the Schr\(\ddot{\text{o}}\)dinger ODE (\ref{radial_schrodinger_p}) is in fact the radial reduction of eigenvalue problem of the operator (\ref{PDE}) examined in Theorem \ref{theorem_GGdagger}, which we derived at the beginning of the chapter, in (\ref{radial_schrodinger}). Numerical investigation shows that the potential corresponding to the Schr\(\ddot{\text{o}}\)dinger equation (\ref{radial_schrodinger_t}) is always above the mass threshold, implying the only bound state solutions are given by solving equation (\ref{radial_schrodinger_p}). Hence we deduce that, in the particular case of radial symmetry, the eigenfunctions of the operator examined in Theorem \ref{theorem_GGdagger} give all the possible shape modes of a radially symmetric \(N\)-vortex.

\section{Conclusions}
In this paper we investigated the internal modes of a general vortex solution in the gauged \(\mathbb{C}P^1\) sigma model with target \(S^2\).  We developed a formalism for computing the Jacobi operator obtained from the second variation of the energy functional which exploits the Bogomol'nyi decomposition. This operator can be factorized into a product two first order operators, one of which is the \(L^2-\)adjoint of the other. To ensure the gauge is fixed, we built a new Jacobi operator which proved to have the same spectrum as the original Jacobi operator. This new operator allows for a much easier computation of the shape modes. In particular, for a general vortex solution on \(\mathbb{R}^2\), we proved the existence of at least one shape mode for $\tau$ close to $0$. We also proved existence of at least one shape mode for all pure North vortices (having empty South divisor $D_-$) for $\tau\in(0,1)$. The construction of these shape modes requires one only to solve a single scalar PDE, the eigenvalue problem of a certain rather simple Schr\(\ddot{\text{o}}\)dinger operator. This is very surprising, as the computation of shape modes generally requires solving the eigenvalue problem for the Jacobi operator itself,  a system of coupled PDEs. We also remarked that this formalism consists of a general approach, and can therefore be applied to other models which admit a Bogomol'nyi decomposition. 

We have used a numerical scheme based on the Newton-Raphson method to obtain the shape mode and its corresponding eigenvalue for a North vortex with topological charge \(N=1,2,3\), and for various values of \(\tau\). Our numerics agreed with the results proven analytically in Section 3, and further revealed that the shape mode for a 1-vortex in this model is generally weakly bound for all values of \(\tau\in(0,1)\).

We remark that, as a by-product of this numerical scheme, we have computed the asymptotic parameter $q(\tau)$ for a single North vortex which, as far as we are aware, has not previously been reported in the literature (except for $\tau=0$ \cite{CP1geometry}). This quantity, which we plot in Figure \ref{q_vs_tau}, may be endowed with physical significance via the point vortex formalism \cite{Speight_particle_interpretation}. The asymptotic fields of the vortex coincide with the solution of the linearization of the model about the vacuum in the presence of a scalar monopole of charge $-q(\tau)$ and a magnetic dipole of moment $-q(\tau)\hat{\bf k}$ at the vortex core (the origin). By an obvious symmetry, the South antivortex corresponds to a point particle with monopole charge $q(-\tau)$ and dipole moment 
$q(-\tau)\hat{\bf k}$. By analyzing the forces between widely separated and slowly moving point particles of this type within the linearized field theory, one can derive a conjectural asymptotic formula for the $L^2$ metric on the moduli space of static $(k_+,k_-)$ vortices, valid in the region where all (anti)vortices are well separated. The result, due to Garcia Lara \cite{Garcia-Lara_coexistence_compact}, is
$$
g\sim 2\pi\sum_{r}(1-\sigma_r\tau)|dz_r|^2
-\frac{1}{4\pi}\sum_{r}\sum_{s\neq r}\sigma_r\sigma_sq(\sigma_r\tau)q(\sigma_s\tau)K_0(\sqrt{1-\tau^2}|z_r-z_s|)|dz_r-dz_s|^2,
$$
where $z_1,z_2,\ldots,z_{k_++k_-}\in\C$ are the positions of the (anti)vortices and $\sigma_r=+1$ for a North vortex and $-1$ for a South antivortex. Having computed $q(\tau)$ we now know these formulae quantitatively. 

A generalised Abelian-Higgs theory has recently been developed in \cite{Yang_generalised_AH} which allows for a Bogomol'nyi decomposition and the coexistence of vortices and anti-vortices for an energy functional with a general potential. It would be interesting to explore the existence of shape modes in these models using the formalism presented in this paper. As pointed out in \cite{Yang_generalised_AH}, one can choose a suitable potential to control the curvature of strings in a cosmological setting. Computation of shape modes in such a scenario could therefore offer much information about the dynamics of cosmic strings in the early universe. 

Another natural question is whether some version of this formalism can be modified to construct shape modes of Chern-Simons deformed vortices, either in the conventional model with target $\C$ \cite{leeleemin} or the model with target $\CP^1$ \cite{kimleelee}. These models still have a Bogomol'nyi decomposition, but it is conceptually more subtle: the solutions are stationary rather than static (they have kinetic energy), and one of the ``Bogomol'nyi" equations is Gauss's law, the Euler-Lagrange equation associated with $A_0$. 
So this extension, if possible at all, is likely to be quite nontrivial.

It would also be interesting to study the shape modes of vortices in a model which further admits some form of integrability. An example would be the hyperbolic vortices in the Abelian-Higgs model on \(\Sigma=\mathbb{D}\) representing the Poincar\'{e} disc model, where solutions can be constructed explicitly using Blaschke functions \cite{Witten_integrable_hypebolic_vortices}. Another possibility would be when \(\Sigma=\mathbb{R}^2\) is endowed with a conformal metric described by a suitable conformal factor leading to an integrable Taubes equation, as observed in \cite{Elizabethan_vortices} for \(N\) coincident vortices. Such properties could allow for a more analytical description of the shape modes, which is not totally reliant on numerical methods.

\subsection*{Acknowledgements}
NG was supported by a UK Engineering and Physical Sciences Research Council (EPSRC) studentship.

\appendix 
\section{Proof of Theorem \ref{thm_taubes_continuity}}\label{appendixA}

We will use the Implicit Function Theorem (IFT) to prove the continuity of the solution in \(\tau\). We first establish the setting by defining a perturbation function which is regular and satisfies its own PDE. Then we define the function to which we will apply the IFT. Before we do this, we first need to check the requirements of the IFT hold true in our context. Hence, our proof is divided into multiple steps intended to check each of the requirements.

For a fixed (and disjoint) divisor pair $(D_+,D_-)$ denote by $(\phivec^\tau,A^\tau)$ the solution of \eqref{B1}, \eqref{B2} with $\phivec^{-1}(\pm\nvec)=D_\pm$. This is known to exist, to be unique up to gauge, to be smooth and, in an appropriate sense, be exponentially spatially localized \cite{Han_general_taubes_existence}. Consider the stereographic projection from the South pole of \(S^2\)
\[u_\tau\coloneq\dfrac{\phi_1^\tau+i\phi_2^\tau}{1+\phi_3^\tau}.\]
From this we easily see
\begin{equation}
\label{bc}
|u_\tau|=\sqrt{\dfrac{1-\phi_3^\tau}{1+\phi_3^\tau}}\to \sqrt{\dfrac{1-\tau}{1+\tau}} \text{ as }|\boldsymbol{x}|\to\infty.\end{equation}
Since $u_\tau$ has the same collections of zeros $D_+$ and poles $D_-$ for all $\tau$ we may obtain $u_\tau$ as a deformation of $u_0$. That is, there exists $h_\tau:\mathbb{R}^2\rightarrow\mathbb{R}$ such that
\begin{equation}\label{deformation_function}
u_\tau=\sqrt\frac{1-\tau}{1+\tau}e^{h_\tau/2}u_0.
\end{equation}
Note that \(h_\tau(\infty)=0\) by \eqref{bc}. This field satisfies the first Bogomol'nyi equation if and only if \cite{garspe} we deform the connexion $A^0$ as
\[A^\tau=A^0-\frac12*dh_\tau,\]
a connexion whose curvature is
\[\star F_\tau=\star F_0+\frac12\Delta h_\tau=\phi_3^0+\frac12\Delta h_\tau.\]
Hence, the deformed pair \((A^\tau,u_\tau)\) satisfies both Bogomol'nyi equations if and only if
\begin{equation}\label{h_PDE}
\Delta h_\tau+2\left(\phi_3^0-\Psi(\tau,h_\tau)\right)=0,
\end{equation}
where
\begin{equation}\label{Phi_formula}
\Psi(\tau,h)=\frac{(1+\tau)(1+\phi_3^0)-(1-\tau)(1-\phi_3^0)e^{h}}{(1+\tau)(1+\phi_3^0)+(1-\tau)(1-\phi_3^0)e^{h}}-\tau
\end{equation}
has been defined so that $\phi_3^\tau-\tau=\Psi(\tau,h_\tau)$. 
We are going to use the fact that \(h_\tau\) satisfies the above PDE to show that $\phi_3^\tau$ depends continuously on $\tau$. Note that $h_0=0$ by construction. \\

\noindent\textbf{Claim 1:} \(\Psi\) is a $C^1$ map \((-1,1)\times H^2(\mathbb{R}^2)\to L^2(\mathbb{R}^2)\).\\

{\it Proof}. By the Sobolev embedding theorem \cite{adams_book_sobolev_spaces}, \(H^{2}(\mathbb{R}^2)\subset C^0_B(\mathbb{R}^2)\), where \(C^0_B(\mathbb{R}^2)\) is the space of all continuous bounded functions on \(\mathbb{R}^2\) which is a Banach space with norm \(||f||_{C^0}=\sup_{\boldsymbol{x}\in\mathbb{R}^2}|f(\boldsymbol{x})|\), and the inclusion map is continuous. Hence there exists \(C>0\) such that, for all \(h\in H^{2}\) we have \(\|h\|_{C^0}\leq C\|h\|_{H^2}\). 

Furthermore, we know \(H^2(\mathbb{R}^2)\) is a Banach algebra \cite{adams_book_sobolev_spaces}. That is, there exists \(C>0\) such that, for all \(h,g\in H^2\), \(hg\in H^2\), and \(\|hg\|_{H^2}\leq C\|h\|_{H^2}\|g\|_{H^2}\).

Note also that \(Q:h\mapsto e^h-1\) is a smooth map \(H^2\to H^2\). To see this, note that it is defined by the power series
\[Q(h)=\sum_{k=1}^\infty \frac{h^k}{k!}\]
which is absolutely convergent on the Banach algebra \(H^2\).

We can simplify \(\Psi\) as follows
\[\Psi(\tau,h)=-\frac{(1-\tau^2)[(e^h-1)-\phi_3^0(e^h+1)]}{(1+\tau)(1+\phi_3^0)+(1-\tau)(1-\phi_3^0)e^{h}}=:\frac{K(h,\tau)}{L(h,\tau)},\]
and partition the plane \(P_+=(\phi_3^0)^{-1}([0,\infty))\), \(P_-=(\phi_3^0)^{-1}(-\infty,0))\). Note that
on \(P_+\)
\[L>(1+\tau)\geq1-|\tau|\]
and on \(P_-\)
\[L>(1-\tau)e^{-\|h\|_{C^0}}\geq(1-|\tau|)e^{-C\|h\|_{H^2}}\]
by the Sobolev embedding theorem. Hence, on the whole plane
\[|\Psi(\tau,h)|\leq (1+|\tau|)e^{C\|h\|_{H^2}}\left[|Q(h)|+(e^{C\|h\|_{H^2}}+1)|\phi_3^0|\right].\]
It follows that \(\Psi(\tau,h)^2\) is integrable, since \(Q(h)\in H^2\subset L^2\) and \(\phi_3^0\) is smooth and exponentially localized. 

Hence $\Psi$ maps $(-1,1)\times L^2 \to H^2$ as claimed. That $\Psi$ is $C^1$ follows from elementary estimates on the difference quotient. \qedsymbol\\

Consider the map 
\[G:(-1,1)\times H^2(\mathbb{R}^2)\to L^2(\mathbb{R}^2)\]
\[G(\tau,h)=\Delta h+2(\phi_3^0-\Psi(\tau,h))\]
so that the PDE defining $h_\tau$ is \(G(\tau,h_\tau)=0\).  $\Delta:H^2\rightarrow L^2$ is a bounded linear map, so it follows immediately from Claim 1 that $G$ is a $C^1$ map.
Define 
\[\bar{G}:H^2(\mathbb{R}^2)\to L^2(\mathbb{R}^2)\]
\[\bar{G}(h)=G(0,h).\]
Then a simple calculation shows 
\[d\bar{G}|_{h=0}=\Delta+1-(\phi_3^0)^2.\]
We will denote this operator $\mathscr{L}:H^2\to L^2$ because it coincides precisely with the Schr\"odinger operator introduced in Theorem \ref{theorem_GGdagger}, in the case $\tau=0$.\\

\noindent\textbf{Claim 2:} The operator \(\mathscr{L}\) is bounded and injective.\\

Boundedness follows trivially from the fact that \(\phi_3^0\) is smooth and bounded. To prove injectivity, we start from \(\mathscr{L}f=0\), multiplying by \(f\) and integrating by parts we obtain 
\[\int^{}_{\mathbb{R}^2}|df|^2d^2x+\int^{}_{\mathbb{R}^2}(1-(\phi_3^0)^2)f^2d^2x=0.\]
Since \(V=1-(\phi_3^0)^2\geq0\), and vanishes only on the finite set $D_+$, we deduce \(f=0\). Therefore \(\mathscr{L}\) is injective. \qedsymbol\\

We next wish to prove that $\L$ is, in fact, invertible. So we need to show that $\L$ is surjective and has bounded inverse. To do this, we will need to show that a certain class of multiplication operators $H^2\to L^2$ are compact. Recall that this means that the operator maps every bounded subset of $H^2$ to a subset of $L^2$ with compact closure. \\

\noindent\textbf{Claim 3:} Let $U:\R^2\to\R$ be a continuous function such that $U=O(|\boldsymbol{x}|^{-2-\epsilon})$ as $|\boldsymbol{x}|\to\infty$ for some $\epsilon>0$.  Then the map $H^2(\R^2)\to L^2(\R^2)$, $f\mapsto Uf$ is compact.\\

\noindent {\it Proof}. We use weighted Sobolev spaces.  For a review, see \cite{Marshall_oxford_fnct_analysis}. Let
\begin{equation}\nonumber
\rho:\mathbb{R}^2\to\mathbb{R},\quad \rho(\boldsymbol{x})=\sqrt{1+|\boldsymbol{x}|^2}.
\end{equation}
For any $\beta\in\R$, let $L^2$, $H^2$, $L^2_{0,\beta}$, $L^2_{2,\beta}$ be the completions of $C^\infty_c(\R^2)$ (the space of smooth functions of compact support) with respect to the following norms:
\begin{align}\nonumber
\|f\|_{L^2}^2 &= \int_{\mathbb{R}^2}f^2\,d^2x\\
\|f\|_{H^2}^2 &= \int_{\mathbb{R}^2}\big(f^2+|\nabla f|^2+|\nabla\nabla f|^2\big)\,d^2x\nonumber\\
\|f\|_{L^2_{0,\beta}}^2 &= \int_{\mathbb{R}^2}\big(\rho^{-\beta}f\big)^2\,\frac{d^2x}{\rho^2}\nonumber\\
\|f\|_{L^2_{2,\beta}}^2 &= \int_{\mathbb{R}^2}\big((\rho^{-\beta}f)^2+|\rho^{1-\beta}\nabla f|^2+|\rho^{2-\beta}\nabla\nabla f|^2\big)\,\frac{d^2x}{\rho^2},\nonumber
\end{align}
in which $\nabla\nabla f$ denotes the Hessian matrix of $f$.   
Theorem 4.18 of \cite{Marshall_oxford_fnct_analysis} tells us that when $\beta<\delta$ the embedding $L^2_{2,\beta}\to L^2_{0,\delta}$ is compact.

The embedding
\begin{equation}\nonumber
H^2\to L^2_{2,1}
\end{equation}
is continuous because $\rho^p\leq 1$ for $p\leq 0$.  The embedding
\begin{equation}\nonumber
L^2_{2,1} \to L^2_{0,1+\epsilon}
\end{equation}
is compact by Theorem 4.18 of \cite{Marshall_oxford_fnct_analysis}.  Our assumptions on $U$ mean that $U\leq C\rho^{-2-\epsilon}$ for some $C>0$.  So the map
\begin{equation}\nonumber
L^2_{0,1+\epsilon}\to L^2_{0,-1}=L^2,\quad f\mapsto Uf
\end{equation}
is continuous.  So the map $H^2\to L^2$, $f\mapsto Uf$ is a composition of a compact map with two bounded linear maps. Therefore it is compact.\qedsymbol \\

\noindent\textbf{Claim 4:} The operator \(\mathscr{L}\) is Fredholm of index zero.\\

\noindent{\it Proof}. Since \(\phi_3^0\) is exponentially decaying as \(|\boldsymbol{x}|\to\infty\), we can apply Claim 4 to $U=-(\phi_3^0)^2$.  Then the map $P:H^2(\mathbb{R}^2)\to L^2(\mathbb{R}^2)$, $f\mapsto -(\phi_3^0)^2f$ is compact.

The operator $\L_0:H^2(\mathbb{R}^2)\to L^2(\mathbb{R}^2)$, \(\L_0=\Delta+1\) is clearly injective. Surjectivity follows via Fourier transforms. Therefore it is Fredholm of index zero. 

So \(\mathscr{L}=\L_0+P\) is equal to the sum of a Fredholm operator \(\L_0\) and a compact operator \(P\), hence it is also Fredholm \cite{Brezis_functional_analysis_ch6} with index
\[ind({\mathscr{L})}=ind(\L_0)=0.\]
\begin{flushright}\qedsymbol\\\end{flushright}

\noindent\textbf{Claim 5:} The operator \(\mathscr{L}\) is bijective, with bounded inverse.\\

\noindent{\it Proof}. Claims 2 and 4 combined show that \(\mathscr{L}\) is bijective. Hence, we know that \(\mathscr{L}\) is a continuous (bounded) linear operator between two Banach spaces which is also bijective. It is known by Corollary 2.7 from \cite{Brezis_functional_analysis_ch2} that the inverse of such an operator is bounded. \qedsymbol\\

\noindent\textbf{Claim 6:} There exists $\epsilon>0$ such that the map
\begin{align*}
(-\epsilon, \epsilon)&\to H^2(\mathbb{R}^2)\\
\tau&\to h_\tau,
\end{align*}
corresponding to the solution of the PDE (\ref{h_PDE}), is a \(C^1\) map.\\

\noindent{\it Proof}. We have just established that $\mathscr{L}=d\bar G_0:H^2\rightarrow L^2$ is invertible, so the claim follows from the Implicit Function Theorem applied to $G$ at $(\tau,h)=(0,0)$.
\qedsymbol\\

\noindent\textbf{Claim 7:} The map
\begin{align*}\Phi:(-\epsilon,\epsilon)&\to L^2(\mathbb{R}^2)\\
\tau&\to \Psi(\tau,h_\tau)=\phi_3^\tau-\tau
\end{align*}
is \(C^1\).\\

Since \(\tau\mapsto h_\tau\) is \(C^1\), and \((\tau,h)\mapsto\Psi(\tau, h)\) given in (\ref{Phi_formula}) is \(C^1\), the composition \(\tau\mapsto\Psi(\tau,h_\tau)=\phi_3^\tau-\tau\) is also \(C^1\). This concludes the proof of the theorem. \qedsymbol \\

\bibliographystyle{amsplain}

\bibliography{Bibliography}

\end{document}